\documentclass[a4paper,fleqn,usenatbib,useAMS]{mnras}

\usepackage{newtxtext,newtxmath}
\usepackage[T1]{fontenc}
\usepackage{ae,aecompl}

\usepackage{graphicx}	% Including figure files
\usepackage{amsmath}	% Advanced maths commands
\usepackage{amssymb}	% Extra maths symbols

\newcommand{\G}{{\it Gaia}}
\newcommand{\Teff}{$T_\mathrm{eff}$}
\newcommand{\Logg}{$\log g$}
\newcommand{\Feh}{$\mathrm{[Fe/H]}$}

\newcommand{\SME}{\textit{SME}}
\newcommand{\kms}{\,km\,s$^{-1}$}

\newcommand{\vsin}{$v \sin i$}

\title[GALAH: emission-line stars]{The GALAH survey: Characterization of emission-line stars with spectral modelling using autoencoders}

\author[K. \v{C}otar et al.]{
Klemen \v{C}otar,$^{1}$\thanks{Contact e-mail: \href{mailto:klemen.cotar@fmf.uni-lj.si}{klemen.cotar@fmf.uni-lj.si}} Toma\v{z} Zwitter,$^{1}$ Gregor Traven,$^{2}$ Joss Bland-Hawthorn,$^{3,4}$ \newauthor
Sven Buder,$^{3,5,6}$ Michael R. Hayden,$^{3,4}$ Janez Kos,$^{1}$ Geraint F. Lewis,$^{4}$ \newauthor
Sarah L. Martell,$^{3,7}$ Thomas Nordlander,$^{3,5}$  Dennis Stello,$^{7}$ Jonathan Horner,$^{8}$ \newauthor Yuan-Sen Ting,$^{5,9,10,11}$ Maru\v{s}a \v{Z}erjal,$^{5}$ and~the~GALAH~collaboration
\\
\\
$^{1}$ Faculty of Mathematics and Physics, University of Ljubljana, Jadranska 19, 1000 Ljubljana, Slovenia \\
$^{2}$ Lund Observatory, Department of Astronomy and Theoretical Physics, Box 43, SE-221 00 Lund, Sweden \\
$^{3}$ ARC Centre of Excellence for All Sky Astrophysics in Three Dimensions (ASTRO-3D), Canberra, ACT 2611, Australia \\
$^{4}$ Sydney Institute for Astronomy, School of Physics, The University of Sydney, Sydney, NSW 2006, Australia\\
$^{5}$ Research School of Astronomy and Astrophysics, The Australian National University, Canberra, ACT 2611, Australia\\
$^{6}$ Max Planck Institute  for Astronomy (MPIA), Koenigstuhl 17, 69117 Heidelberg, Germany\\
$^{7}$ School of Physics, University of New South Wales, Sydney, NSW 2052, Australia\\
$^{8}$ Centre for Astrophysics, University of Southern Queensland, Toowoomba, QLD 4350, Australia\\
$^{9}$ Institute for Advanced Study, Princeton, NJ 08540, USA\\
$^{10}$ Department of Astrophysical Sciences, Princeton University, Princeton, NJ 08544, USA\\
$^{11}$ Observatories of the Carnegie Institution of Washington, 813 Santa Barbara Street, Pasadena, CA 91101, USA
}

\date{Accepted XXX. Received YYY; in original form ZZZ}

\pubyear{2020}

\begin{document}
\label{firstpage}
\pagerange{\pageref{firstpage}--\pageref{lastpage}}
\maketitle

\begin{abstract}
We present a neural network autoencoder structure that is able to extract essential latent spectral features from observed spectra and then reconstruct a spectrum from those features. Because of the training with a set of unpeculiar spectra, the network is able to reproduce a spectrum of high signal-to-noise ratio that does not show any spectral peculiarities even if they are present in an observed spectrum. Spectra generated in this manner were used to identify various emission features among spectra acquired by multiple surveys using the HERMES spectrograph at the Anglo-Australian telescope. Emission features were identified by a direct comparison of the observed and generated spectra. Using the described comparison procedure, we discovered $10,364$ candidate spectra with a varying degree of H$\alpha$/H$\beta$ emission component produced by different physical mechanisms. A fraction of those spectra belongs to the repeated observation that shows temporal variability in their emission profile. Among emission spectra, we find objects that feature contributions of a nearby rarefied gas (identified through the emission of [NII] and [SII] lines) that was identified in $4004$ spectra, which were not all identified as having H$\alpha$ emission. Positions of identified emission-line objects coincide with multiple known regions that harbour young stars. Similarly, detected nebular emission spectra coincide with visually-prominent nebular clouds observable in the red all-sky photographic composites.
\end{abstract}


%%%%%%%%%%%%%%%%%%%%%%%%%%%%%%%%%%%%%%%%%%%%%%%%%%

%%%%%%%%%%%%%%%%% BODY OF PAPER %%%%%%%%%%%%%%%%%%

\section{Introduction}
The identification of peculiar stars, whose spectra contain emission lines, is of interest to a wide field of stellar research. Spectral complexity of such stars brings insight into the ongoing physical processes on and around the star. Emission features in stellar spectra might adversely impact the quality of stellar parameters and abundances determined by automatic data analysis pipelines that are configured to produce the best results for most common stellar types. Examples of how these features might compromise spectroscopic measurements when we assume that a star is not peculiar include the determination of effective temperature \citep{2011A&A...531A..83C, 2018A&A...615A.139A, 2019A&A...624A..10G}, computation of stellar mass \citep{2016ApJ...823..114N, 2016A&A...594A.120B}, and the effects of self broadening on line wing formation \citep{2000A&A...363.1091B, 2008A&A...480..581A}. Highly accurate measurement of the hydrogen absorption profiles are needed in those cases. Any deviations in the line shapes from model predictions would produce misleading results. We would therefore like to know if the investigated line is modified by additional, unmodelled physical process or spectral reduction process.
\textit{}
Stars with evident emission lines populate a wide variety of regions on the HR diagram. Because of possible overlaps between different stellar types, detailed photometric (especially in the infrared region where warm circumstellar dust disc can be identified) and spectroscopic observations are needed for an accurate physical explanation of the observed features. An examples of such work is presented in \citet{2019MNRAS.488.5536M}, who performed detailed a multi-band photometric study of an emission-line star, originally discovered on objective prism plates. The detailed photometric time-series study described in that work, together with observations of the star's infrared excess, led to the star VES~263 being identified as a massive pre-main-sequence star and not a semi-regular AGB cool giant as classified previously. In a similar way, \citet{2020arXiv200207852L} performed an analysis of the stellar object V* CN Cha, which had previously been identified as an emission star. By studying a long photometric time-series of the star, they concluded that the object was most likely a symbiotic binary star system whose emission was lined to a long-duration, low-luminosity nova phase.

Numerous different physical processes that can contribute to the complex shapes of the H$\alpha$ emission profile are given by \citet{1996A&AS..120..229R, 2011AJ....141..150J, 2014ApJ...795...82S, 2018AJ....156...97I}, who compare observations with expected physical models. Following the classification scheme introduced by \citet{2007ASSL..342.....K}, emission-line stars are predominately observed in close binaries, earliest-type, latest-type, and pre-main sequence stars. For systems in which mass accretion is occurring, the examination of emission lines can allow the mass accretion rate onto the central star to be estimated \citep{2003ApJ...582.1109W, 2004A&A...424..603N}. The procedure involves measuring simple indices (such as the equivalent width and broadening velocity) of the emission lines in the stars's spectrum.

In recent years, multiple dedicated photometric and spectroscopic surveys \citep[e.g.][]{2008MNRAS.384.1277W, 2008MNRAS.388.1879M, 2012ApJS..200...14M, 2012AJ....143...61N, 2014MNRAS.440.2036D, 2016MNRAS.456.1424A, 2016ASPC..505...66N}, and exploratory spectral classifications of large unbiased all-sky spectroscopic observational datasets \citep[e.g.][]{1999A&AS..134..255K, 2012MNRAS.425..355R, 2015A&A...581A..52T, 2016ASPC..505...66N, 2016RAA....16..138H, 2017ApJS..228...24T} have been performed, each finding from hundreds to tens of thousands of interesting emission-line stars. Some of these surveys provide a basic physical classification in addition to an emission detection. Therefore they can be used as source lists for further in-depth studies of individual stars.

If a star is engulfed in a hot rarefied interstellar medium or stellar envelope, emission features of the forbidden lines (the most commonly studied of which are the [NII] and [SII] lines) could be observed in its spectrum, providing an insight into the temperature, density, intrinsic movement, and structure of its surrounding interstellar environment \citep{1973ApJ...184...93B, 1993Ap&SS.204..205R, 2005MNRAS.361..813E, 2016A&A...591A..74D, 2017A&A...604A.135D}.

Focusing on spectroscopic data, procedures for the detection of emission lines can roughly be separated into two categories. Simpler procedures searching for obvious emitters above the global continuum \citep[][]{2015A&A...581A..52T, 2016ASPC..505...66N, 2016RAA....16..138H, 2016ASPC..505...66N} and more complex procedures, where the observed spectrum is compared to an expected stellar spectrum of a normal star \citep[][]{2013ApJ...776..127Z}. The reference spectra in the latter case can be generated using exact physics-based stellar modelling or data-driven approaches. Of these, the data-driven approaches can be separated into supervised and unsupervised generative models, where, for the later, it is not required to provide an estimate of the stellar parameters for a given spectrum in advance. To predict a reliable model using supervised models, we must determine the correct stellar labels of an emission star in advance. This can pose a serious limitation if the strongest lines in the acquired spectrum can be populated by an emission feature, which happens for \G\ and RAVE spectra \citep{2013ApJ...776..127Z}. In light of the future publication of Gaia RVS spectra as part of Gaia DR3 for hundreds of million of stars, it is thus important to develop tools to identify emission-line stars, as we aim to do in this study via GALAH spectra.
% kak odstavek: poudariš da boš tukaj delal to for in zakaj so autoencoderji kul, recimo

The paper is structured in the following order. We begin with the description of the used spectroscopic data in Section \ref{sec:data_galah}. In Section \ref{sec:det_chart} we explain our analysis pipeline whose main components are the generation of reference spectra (Section \ref{sec:ref_modeling}) and identification of multiple emission features (Sections \ref{sec:hahbemis} and \ref{sec:nebularemis}). The temporal variability of detected emissions is analysed in Section \ref{sec:temporal}. The results are discussed and summarised and discussed in Section \ref{sec:discussion}.

%%%%%%%%%%%%%%%%%%%%%%%%%%%%%%%%%%%%%%%%%%%%%%%%%%

\section{Data}
\label{sec:data_galah}
The spectroscopic data used in this study was taken from the main GALactic Archaeology with HERMES (GALAH) survey \citep{2015MNRAS.449.2604D}, the K2-HERMES survey \citep{2018AJ....155...84W}, the TESS-HERMES survey \citep{2018MNRAS.473.2004S}, and the dedicated HERMES open clusters survey (De Silva et al. in preparation) and the HERMES Orion star forming region (Kos et al. in preparation) survey. All of the spectra were acquired by the High Efficiency and Resolution Multi-Element Spectrograph \citep[HERMES,][]{2010SPIE.7735E..09B, 2015JATIS...1c5002S}, a multi-fibre spectrograph mounted on the $3.9$-metre Anglo-Australian Telescope (AAT) at the Siding Spring Observatory, Australia. The spectrograph has a resolving power of R $\sim 28,000$ across four wavelength ranges (4713~--~4903~\AA, 5648~--~5873~\AA, 6478~--~6737~\AA, and 7585~--~7887~\AA), also referred to as blue, green, red, and infrared spectral arms. Of the four, we use only the first (blue) and the third (red) in our study, as they cover the wavelength regions where interesting Balmer and forbidden emission lines can be seen and detected.

The combined dataset consist of $669,845$ successfully reduced stellar spectra, of which a small fraction are repeated observations. All acquired spectra were homogeneously reduced to one dimensional spectra, continuum normalised, and shifted to the stellar reference frame \citep[a detailed description of the algorithm used can be found in][]{2017MNRAS.464.1259K}. All surveys combined include more spectra than the main GALAH survey alone, but at the same time break the simple rule, adhered to in that main survey, of a simple magnitude limited selection function (Sharma et al. in preparation) that is desired for population studies and comparison with synthetic galactic models. The exact selection function is not important in our case as we are not performing any population studies, but are only trying to find as many emission-line objects as possible.

\begin{figure}
	\centering
	\includegraphics[width=\columnwidth]{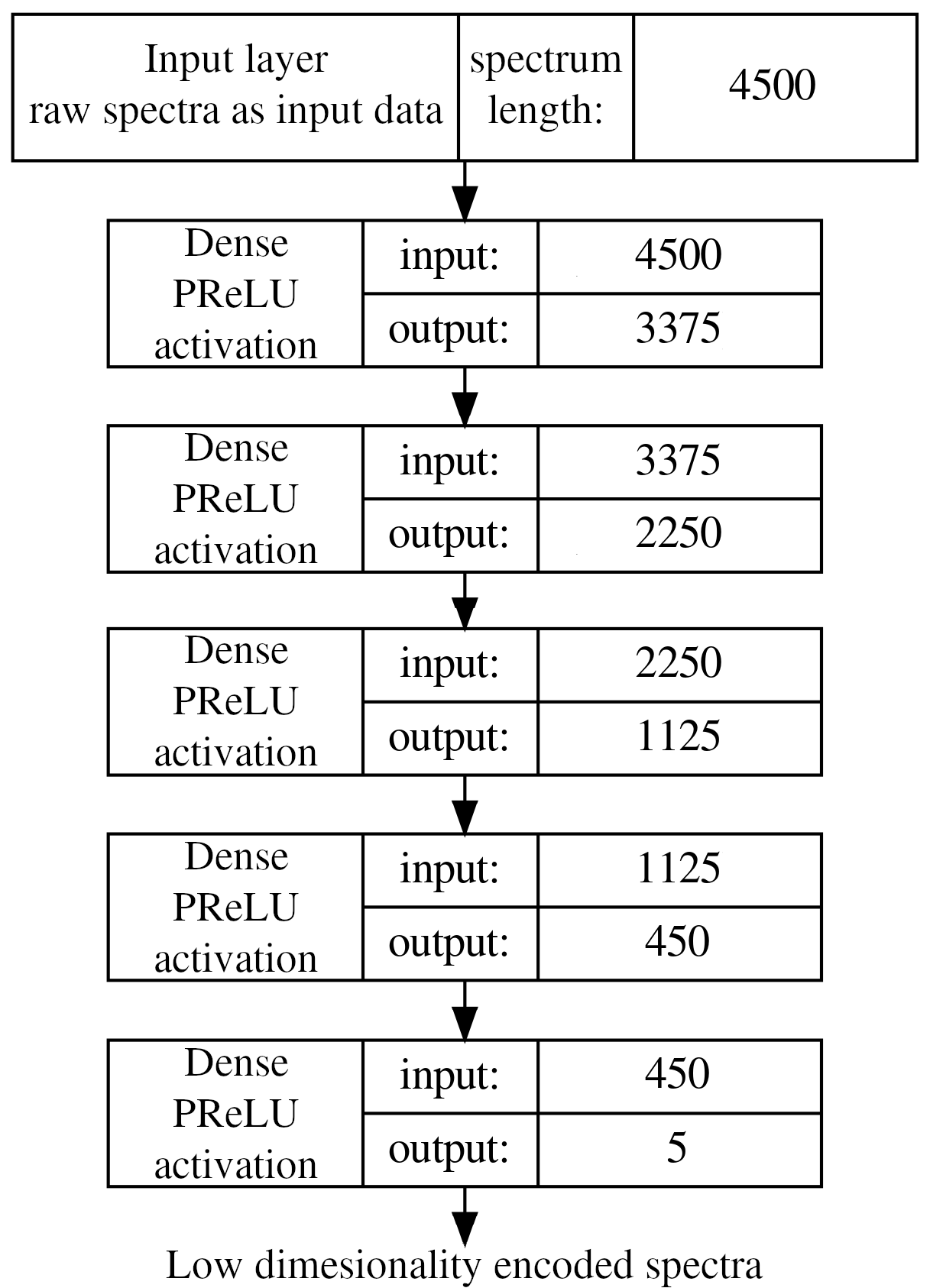}
	\caption{Visual representation of an encoder part of the used autoencoder structure for the red spectral arm. After the input spectra are encoded, they are passed through the same inverted architecture to produce modeled low-noise spectra. The value in the right most column indicates a number of input and output connections to neighboring layers. The number of nodes in a layer is equal to the output value. The input spectrum length is given as the number of wavelength bins in a spectrum.}
	\label{fig:autoann}
\end{figure}

The stellar atmospheric parameters and individual abundances derived from our normalised spectra were analysed with the same adaptation of the Spectroscopy Made Easy \citep[\SME,][]{1996A&AS..118..595V, 2017A&A...597A..16P} software that is described in-depth by Buder et al. (in preparation) as part of the latest GALAH data release (DR3) that includes fully reduced spectra and derived parameters.

Our algorithm for the detection of emission-line spectra, described in detail below, uses the normalised GALAH spectra that were already corrected for telluric absorptions and had sky spectral emission contributions removed. The correct sky removal (described in more detail in Section \ref{sec:skyemis}) is essential as one of the telluric lines falls inside the range of the H$\alpha$ line. 

%%%%%%%%%%%%%%%%%%%%%%%%%%%%%%%%%%%%%%%%%%%%%%%%%%

\section{Detection and characterization}
\label{sec:det_chart}
The first attempts to discover H$\alpha$/H$\beta$ emission spectra in GALAH survey observations were performed by \citet{2017ApJS..228...24T}, who use the unsupervised dimensionality reduction technique t-SNE \citep{2013arXiv1301.3342V} to group morphologically similar spectra. As the amplitude and shape of the observed emission can vary substantially depending on the astrophysical source, \citet{2017ApJS..228...24T} presumably detected only a portion of the strongest emitters. One of the reasons for this is the manual classification of data clumps determined by the clustering algorithm. In the case of weak emissions in an investigated clump (performed manually by the operator), an expressed emission feature must be strong enough to be visually perceived when looking at a spectrum. To broaden the range of detectability to include spectra with marginal levels of emission as well, a more sophisticated and partially supervised procedure must be employed.

To expand the search, our methodology uses additional prior knowledge about the expected wavelength locations of interesting emission spectral lines. The prior wavelengths are used to narrow down the interesting wavelength regions during the comparison between the spectrum of a possibly peculiar star and an expected (reference) spectrum of a star with similar physical parameters and composition.

\begin{figure}
	\centering
	\includegraphics[width=\columnwidth]{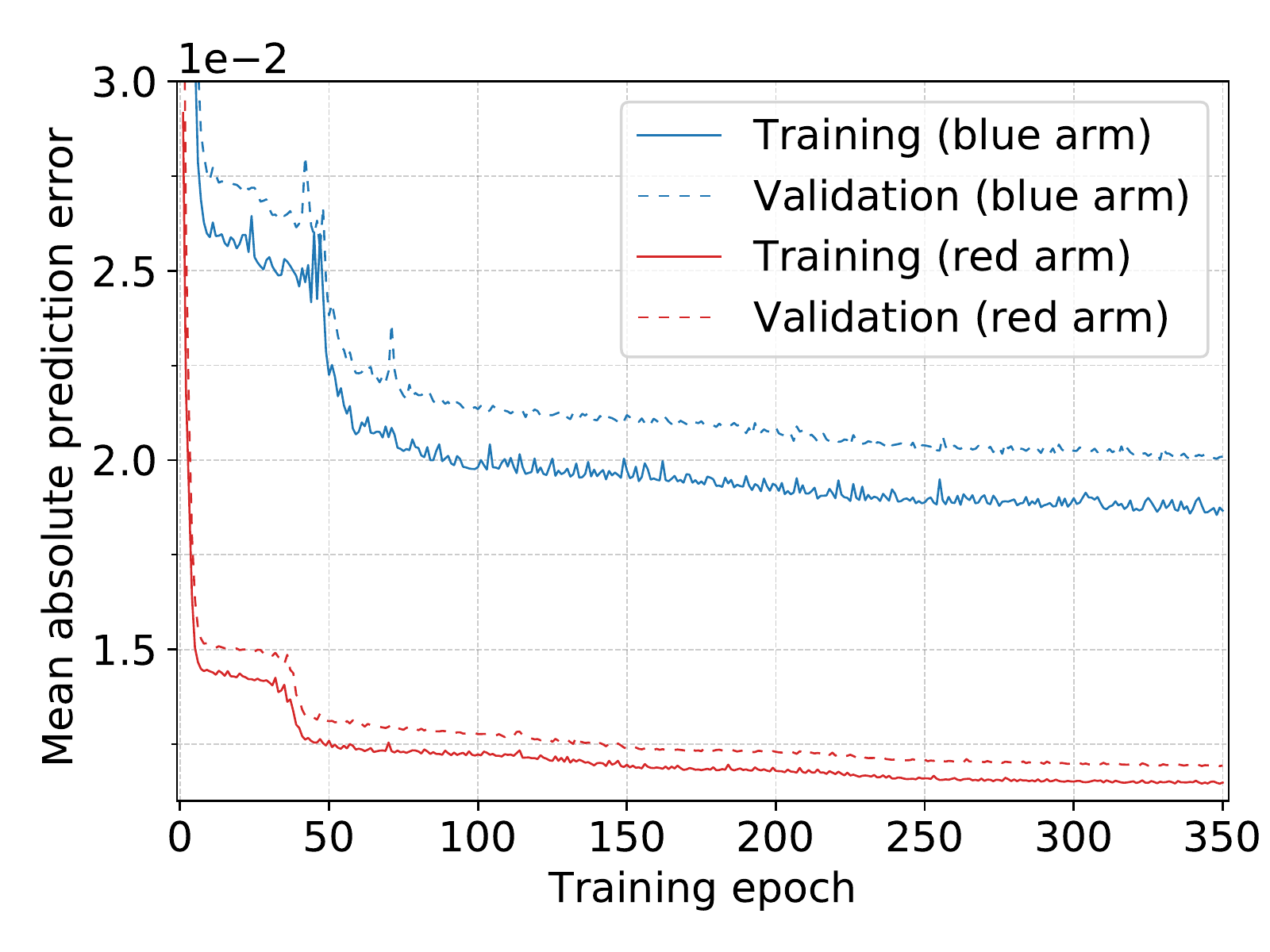}
	\caption{Prediction accuracy of the blue and red arm autoencoders at different training epochs. The prediction error is computed as a sum of all absolute differences between the input and output data set (see Equation \ref{equ_mae}). Shown are training (solid line) and validation curves (dashed line) which do not show any strong model over-fitting on the training set. The curves indicate that both autoencoders learned in a similar way because the same optimiser was used. The blue arm model has a bit higher loss and shows slower learning because of a greater spectral complexity and lower signal to noise ratio in that wavelength region.}
	\label{fig:trainann}
\end{figure}

\begin{figure*}
	\centering
	\includegraphics[width=\textwidth]{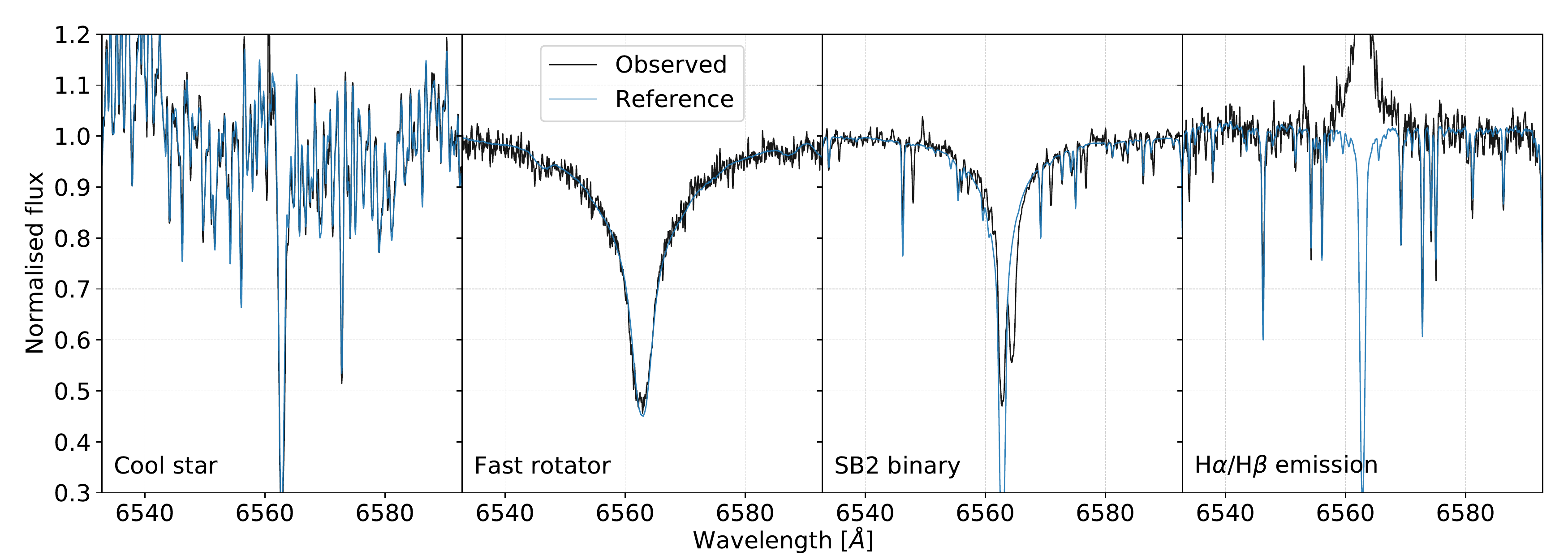}
	\caption{Showing a diversity of spectra that must be processed by our reference spectrum generation scheme. Panels show spectra of the following normal and peculiar stars: cool, hot fast-rotating, spectroscopic binary, and H$\alpha$/H$\beta$ emission star. All examples show that the autoencoder network did reproduce the observed shapes of the normal spectra (first two) and not the peculiar spectra (last two) as desired from the reference spectrum generator. The original spectra are shown in black and reconstructed in blue.}
	\label{fig:refann}
\end{figure*}

\subsection{Spectral modelling using autoencoders}
\label{sec:ref_modeling}
A reference or a synthetic spectrum of a normal, non-emissive star, can be produced by a multitude of physically-based computational stellar models \citep{1993sssp.book.....K, 2005A&A...442.1127M, 2012ASInC...6...53D} or supervised generative data-driven approaches \citep{2015AAS...22530207N, 2019ApJ...879...69T}, whose common weakness is the need for prior knowledge of at least an approximate stellar parameters of an analysed stars used by the data-driven algorithm.

As some of our spectra do not have determined stellar parameters or they are flagged with warning signs that indicate different reduction and analysis problems (missing infrared arm, various reduction issues, bad astrometric solutions, \SME\ did not converge etc.), we focused on an unsupervised spectral modelling to produce our set of reference spectra. Given the large size of available training data set, we chose to use an autoencoder type of an artificial neural network (ANN) that is rarely used to analyse astronomical data. Its current use ranges from data denoising \citep{2017ChA&A..41..282Q, 2019arXiv190303105S, 2019MNRAS.485.2628L} to unsupervised feature extraction and feature based classification \citep{2015MNRAS.452..158Y, 2017RAA....17...36L, 2017ChA&A..41..318P, 2018arXiv180901434K, 2019arXiv191104320C, 2019ApJS..240...34M, 2019PASP..131j8011R}.

An autoencoder is a special kind of ANN, shaped like an hourglass, that takes input data (a stellar spectrum in our case), reduces it to a selected number of latent features (a procedure known as encoding) and tries to recover the original data from the extracted latent features (decoding process). Our dense, fully connected autoencoder consists of the data input layer, four encoding layers, a middle feature layer, four decoding layers and the output layer. The number of nodes (or latent spectral features) in the encoding part slowly decreases in the following arbitrary selected order: $75\%$, $50\%$, $25\%$, and $10\%$ of input spectral wavelength samples (4500 in the case of the red spectral arm). The exact numbers of nodes at each layer are shown in Figure \ref{fig:autoann}. At the middle feature layer, the autoencoder structure reduces to only $5$ relevant extracted features. Selecting a higher number of extracted features would also mean that the ANN structure could extract more uncommon spectral peculiarities which is not what we want. In our case, the goal is the reconstruction of a normal non-peculiar spectrum by extraction of a few relevant spectral features. Additionally, because of the low number of extracted features, our decoded output spectrum is a smoothed and denoised version of an input spectrum.

A visual representation of the described architecture is shown in Figure \ref{fig:autoann}. The shape of the decoding structure of the autoencoder is the same, except in a reverse order. The Parametric Rectified Linear Unit \citep[PReLU,][]{2015arXiv150201852H} activation function defined as
\begin{equation}
	f(x) = \begin{cases}
	x, & \mathrm{if} x > 0 \\
	ax, & \mathrm{if} x \le 0
	\end{cases}
	\label{equ:prleu}
\end{equation}
is used for all nodes of the network, with the exception of the final output layer that uses a linear (i.e. identity) activation function. The $x$ denotes one spectrum flux value in the first layer and one latent feature in the remaining layers. The free parameter $a$ in Equation \ref{equ:prleu} is optimised during the training phase.

If the network learns a physics-based generative model of a stellar spectrum, information contained in the extracted features should be related to real physical parameters, such as \Teff, \Logg, \Feh, and \vsin, or their mathematical combinations. 

\begin{figure*}
	\centering
	\includegraphics[width=\textwidth]{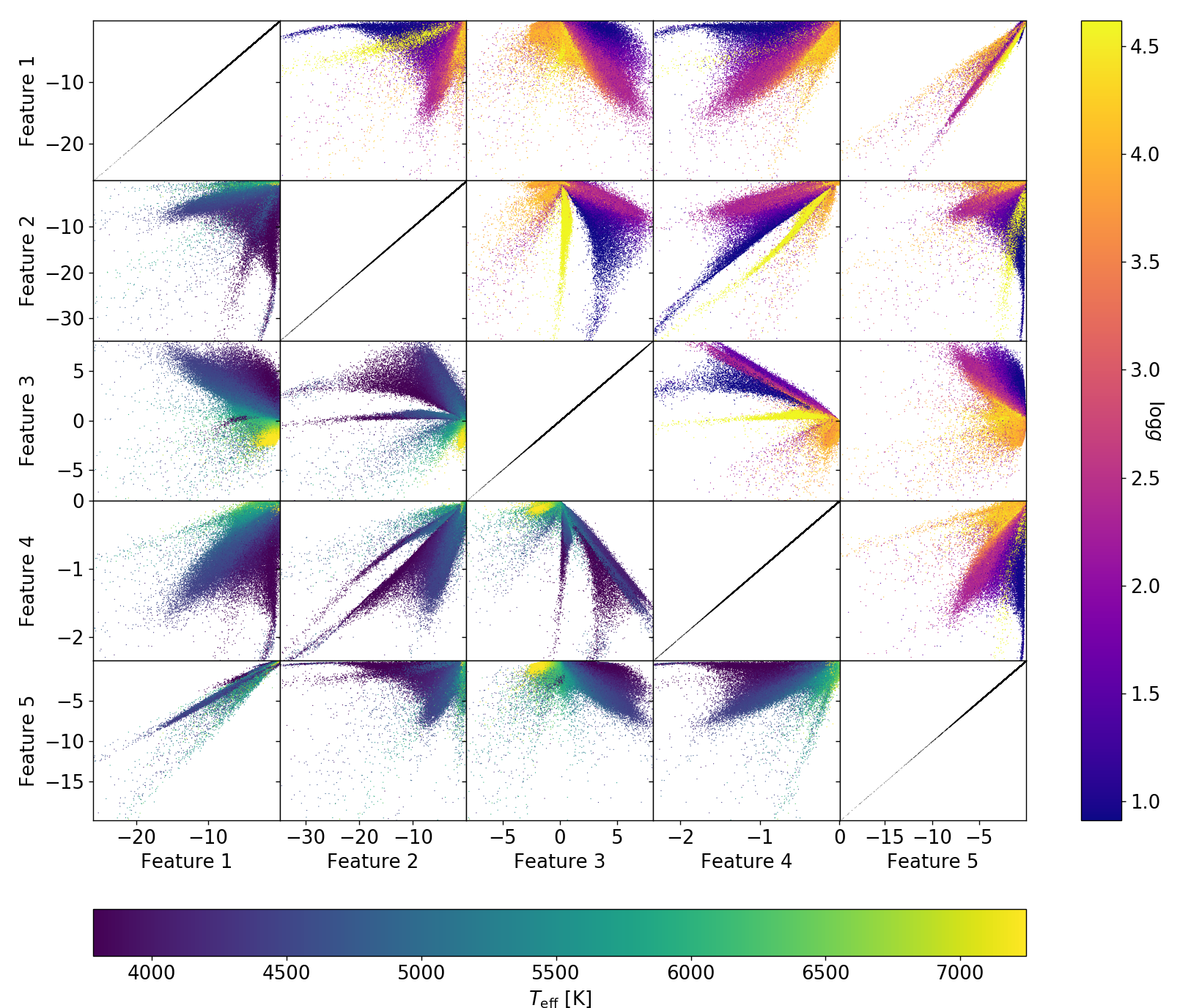}
	\caption{Correlation between extracted latent features and physical parameters. Scatter plots between different features are colored by the GALAH physical parameters of original spectra. Points in the lower triangle are colored by their \Teff\ and in upper triangle by their \Logg. Associated colour mappings are given below the figure (for the lower triangle) and on its right side (for the upper triangle). Presented are results for the red arm autoencoder.}
	\label{fig:latent_ccd3_1}
\end{figure*}

To train our autoencoder, we created a set of presumably normal spectra (with no emission features), resampled to a common wavelength grid ($\delta \lambda$ equal to $0.04$ and $0.06$~\AA\ for the blue and red arm) whose coverage is slightly wider than the range of an individual HERMES arm to account for variations in wavelength span because of radial velocity. Observations that did not completely fill the selected range were padded with continuum value of 1. To be classified as normal, spectra must suffice the following selection rules: signal to noise ratio (SNR) in the green arm must be greater than $30$, a spectrum must not contain any known reduction issues \citep[\texttt{red\_flag}~=~0 in][]{2017MNRAS.464.1259K} and have valid spectral parameters \citep[\texttt{flag\_sp}~<~16 in Buder et al. in preparation][]{}. Although choosing \texttt{flag\_sp}~=~0 returns the spectra with the most trustworthy parameters, we choose to use this higher cutoff in \texttt{flag\_sp} to filter out only the strangest spectra and not to produce a set of spectra with well defined parameters. Spectra with 0~<~\texttt{flag\_sp}~<~16 include objects with bad astrometric solution, unreliable broadening, and low SNR that are still useful for our training process. From \citet{2017ApJS..228...24T, 2018MNRAS.478.4513B} and \citet{2019MNRAS.483.3196C}, we know that some GALAH spectra display peculiar chemical composition or consist of multiple stellar components, therefore we removed all identified classes with the exception of stars classified as hot or cold that are actually treated as normal spectra in our case. Even such a rigorous filtering approach can miss some strange spectra.

After we applied these quality cuts, we were left with $482,900$ spectra, of which last $10\%$ were used as an independent validation set during the training process. Before the training, normalised spectra were inverted ($1$~$-$~normalised flux), which sets the continuum level to a value of $0$. The inversion improved the model stability and decreased the required number of training epochs.

The described autoencoder was trained with the Adam optimisation algorithm \citep{2014arXiv1412.6980K} for $350$ epochs. At every epoch all training spectra were divided into multiple batches of $40,000$ spectra, whose content is randomised at every epoch. A batch is a subset of data that is independently used during a training process. Such splitting and randomisation of training spectra into batches decreases the probability of model over-fitting. To enable the selection of the best network model, it was saved after the end of every training epoch. 

The loss score minimised by the Adam optimiser, shown in Figure \ref{fig:trainann}, was computed as a mean absolute error (MAE) between the input observed and decoded spectra defined as:
\begin{equation}
\label{equ_mae}
	loss_\mathrm{MAE} = \frac{1}{N n_\lambda} \sum_{n=1}^{N}\sum_{i=1}^{n_\lambda}\left|f_{\mathrm{ae}, n, i} - f_{\mathrm{obs}, n, i}\right|,
\end{equation}
where N represent the number of all spectra, $n_\lambda$ the number of wavelength bins in each spectrum,  $f_{\mathrm{ae}, n, i}$ the flux value of a decoded spectrum at one of the training epochs, and $f_{\mathrm{obs}, n, i}$ the flux value of a normalised observed spectrum. Such a loss function gives lower weight to gross outliers in comparison to the mean squared error (MSE). At the same time, outputs are closer to a median spectrum of spectra with a similar appearance and less affected by remaining peculiar spectra in the training set. 

After examining the decoded outputs at different epochs in comparison with known normal and peculiar spectra, we decided to use the model produced after $150$ training epochs. After that, overall improvements of the model are minor, which increases the model opportunity to over-fit on a low number of peculiar spectra. After closer inspection of the last epoch, we found indications of over-fitting on known emission stars, which further confirms the validity of choosing less longer trained model (with greater prediction loss) and rejects the need for a longer model training.

To decrease the complexity of a dense neural network and reduce the required training time, two independent autoencoders were trained, separately for the blue and red HERMES spectral arms.

After the training and model selection were completed, all available $669,845$ spectra were run through the same autoencoder to produce their high SNR reference spectra. An example of four such spectra is shown in Figure \ref{fig:refann}.

\subsection{Latent features}
To test the idea of extracted scalar latent features being connected to physical parameters, and to inspect how an autoencoder structure actually orders spectra, we colour coded values of latent features by unflagged physical parameters of input the GALAH spectra. Latent feature scatter plots, colour coded by a different combination of stellar parameters, are presented in Figures \ref{fig:latent_ccd3_1} (with \Teff\ and \Logg\ for the red arm) as well as \ref{fig:latent_ccd1_1} (with \Teff\ and \Logg\ for the blue arm) and \ref{fig:latent_ccd1_2} (with \Teff\ and \Feh\ for the blue arm).

As expected, all plots show colour gradients induced by the changing value of investigated physical parameter. This gives us a confirmation that the derived stellar physical parameters are spectroscopically meaningful and have the strongest influence on the appearance of acquired spectra. Rough physical parameters of previously unanalysed or peculiar spectra can therefore be acquired by averaging the parameter values of their neighborhood in the latent space. Similar procedures for parameter estimation have already been successfully explored by \citet{2015MNRAS.452..158Y, 2017ChA&A..41..318P, 2017RAA....17...36L}.

\begin{figure}
	\centering
	\includegraphics[width=\columnwidth]{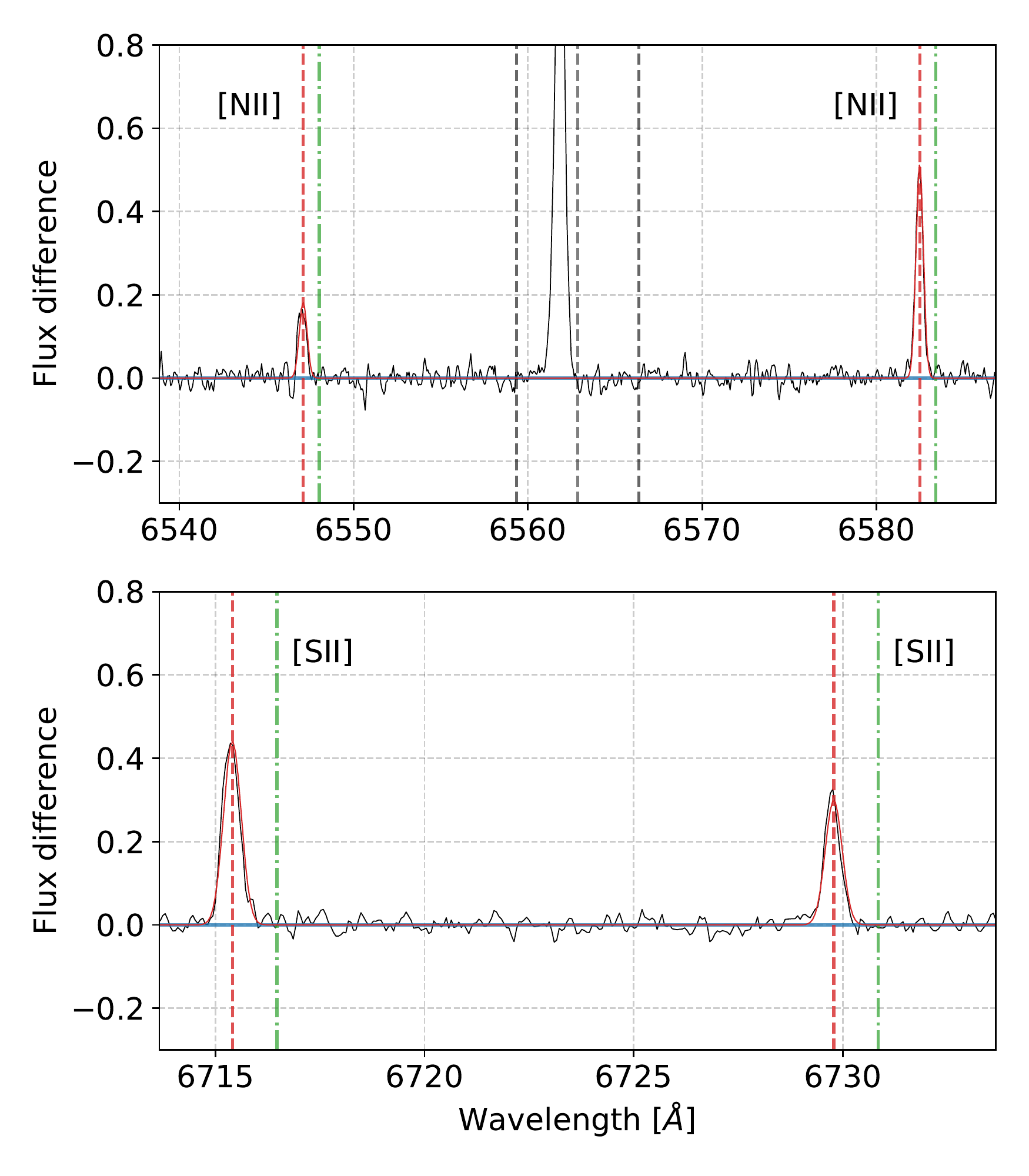}
	\caption{Panels show two different wavelength regions of $f_\mathrm{diff}$ for the same star. The top panel is focused on the H$\alpha$ and [NII] nebular lines, while the second panel focuses on [SII] lines. Rest wavelengths of both nebular doublets are given by the green dash-dotted vertical lines. Their fitted locations, affected by a gas cloud movement, are given by the red dashed vertical lines. The EW(H$\alpha$) integration range is bounded by the central black dashed vertical lines on the top panel.}
	\label{fig:emissfit}
\end{figure}

\subsection{H$\alpha$ and H$\beta$ emission characterization}
\label{sec:hahbemis}
The detection of emission components in spectra is based on a spectral difference $f_\mathrm{diff}$, computed as:
\begin{equation}
	\label{equ:spec_diff}
	f_\mathrm{diff} = f_\mathrm{obs} - f_\mathrm{ref},
\end{equation}
where $f_\mathrm{obs}$ and $f_\mathrm{ref}$ are the observed spectrum and the generated reference spectrum respectively. The result of a computed difference $f_\mathrm{diff}$ for an emission spectrum is shown in the top panel of Figure \ref{fig:emissfit}. Ideally, this computation would enhance only mismatch between both spectra, with inclusion of spectral noise, if both represent a star with the same stellar physical parameters. During the initial processing, we found out that some observed spectra have slight normalisation problems, therefore we re-normalised them prior to difference computation. As the targeted reference spectrum $f_\mathrm{ref}$ is known and has a continuum level close to a median value of similar stars in the training set, we first compute a spectral ratio $f_\mathrm{div}$, defined as:
\begin{equation}
	\label{equ:spec_div}
	f_\mathrm{div} = \frac{f_\mathrm{obs}}{f_\mathrm{ref}}.
\end{equation}
The resulting ratio can be viewed as a proxy for a renormalisation curve that would bring $f_\mathrm{obs}$ to the same continuum level as $f_\mathrm{ref}$, but would at the same time cancel out any spectral differences between them. To avoid the later, we fited $f_\mathrm{div}$ with a $3^{rd}$ degree polynomial with a symmetrical $2$-sigma clipping, run for five iterations. We used the polynomial fit to renormalise $f_\mathrm{obs}$.

\begin{figure}
	\centering
	\includegraphics[width=\columnwidth]{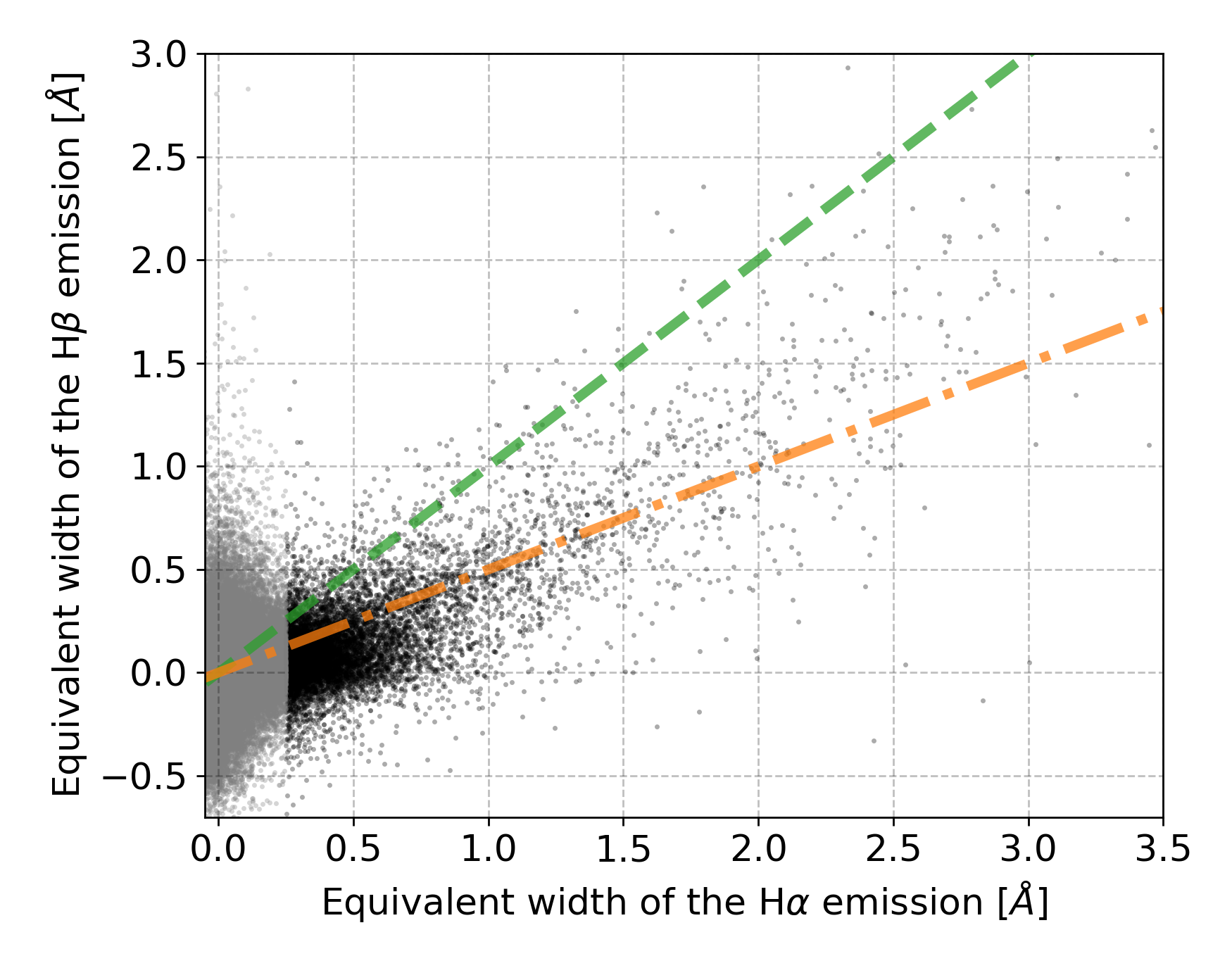}
	\caption{Correlation between equivalent widths of the H$\alpha$ and H$\beta$ emission components for our set of detected stars (defined as having \texttt{Ha\_EW}~>~$0.25$~\AA) as black points. The remaining set of objects is shown with gray dots. All flagged objects and possible spectroscopic binaries are taken out for this plot. The green dashed linear line represent the one-to-one relation and the orange dash-dotted line identicates cases where the equivalent width of the H$\beta$ is half of the H$\alpha$ line.}
	\label{fig:hab_EW}
\end{figure}

To get the first identification of an emission features, we calculate the equivalent width (EW) of the spectral difference in a $\pm3.5$ \AA\ range around the investigated Balmer H$\alpha$ and H$\beta$ lines. The selected range (shown in Figure \ref{fig:emissfit}) is wide enough to encompass emission profile of all spectra, with the exception of a few ones, which have very broad and structured profiles. We kept the width narrow to reduce the effect of spectral noise and nearby sky emission lines (see Section \ref{sec:skyemis}). The correlation between measurements of both equivalent widths is shown in Figure \ref{fig:hab_EW} from which it is evident that the H$\beta$ emission feature is not as strong as the H$\alpha$ feature, but comparable to it.

Alongside the equivalent widths of the residual components (EW(H$\alpha$) and EW(H$\beta$)), we also measured two additional properties of these lines, which give some insight into physical understanding of emission source. The broadening velocity of a line is described by its width at the $10$\% of the line peak (W10\%(H$\alpha$) and W10\%(H$\beta$)) expressed in \kms. The automatic measurement procedure first finds the highest point inside the integration wavelength range and then slides down on both sides of the peak until its strength drops below 10\% of the peak flux value. The broadening velocity is defined as a width between those two limiting cuts. As the computation is done for every object in an unsupervised way, the results are meaningful only for the spectra with evident emission lines. In the case when a low broadening velocity is estimated (equivalent to a very narrow peak), the highest peak could be a residual sky emission line or a cosmic ray streak. By combining EW(H$\alpha$) and W10\%(H$\alpha$), mass accretion could be estimated if emission is of a chromospheric origin.

\begin{figure}
	\centering
	\includegraphics[width=\columnwidth]{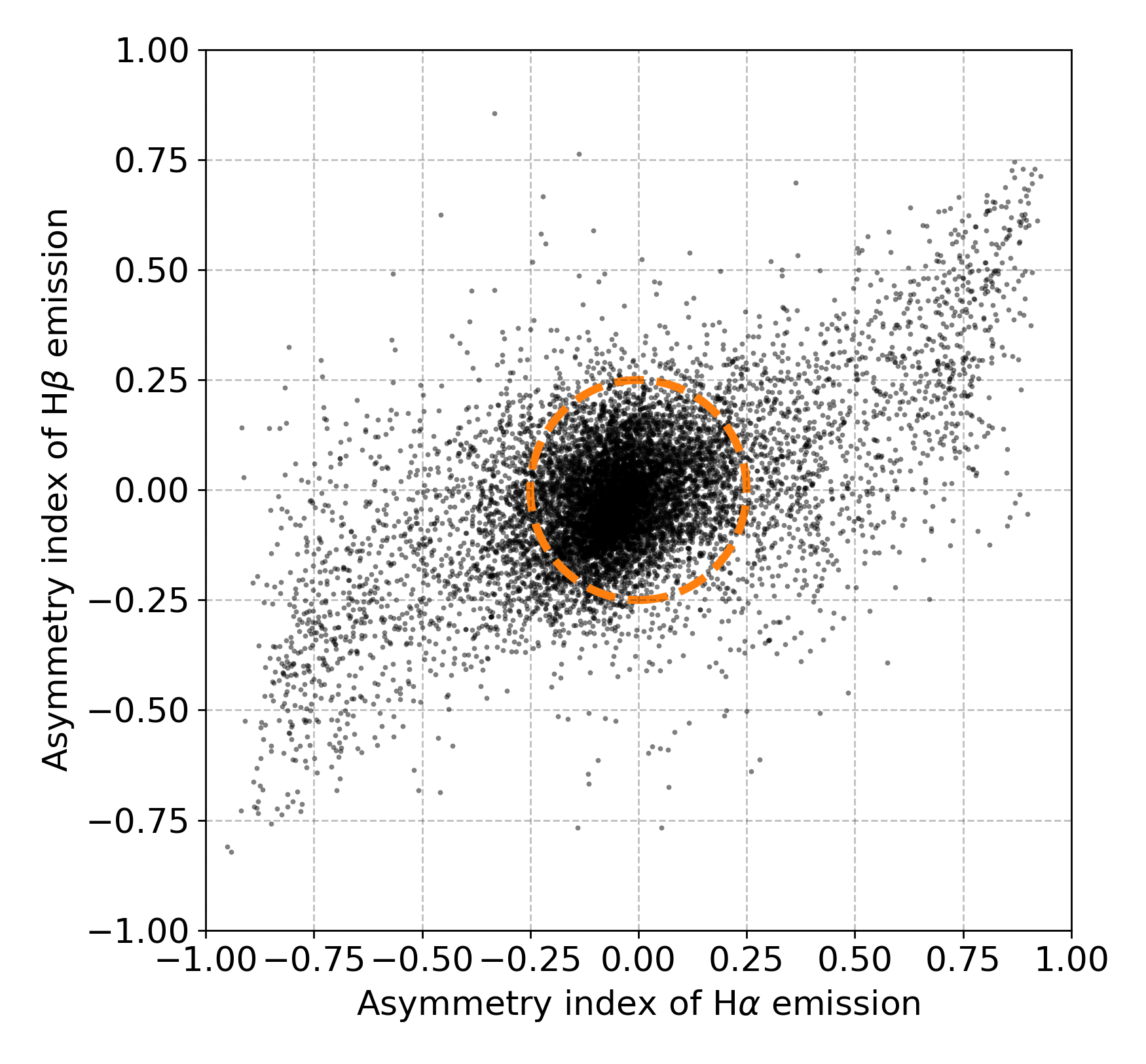}
	\caption{Asymmetry index of objects with prominent emission lines in the integration range around investigated Hydrogen Balmer lines. Objects with index inside the green dashed circle are considered to have a symmetric emission contribution, which can be attributed to a chromospheric activity. Central circular region has a radius of asymmetric index $0.25$.}
	\label{fig:hab_asym}
\end{figure}

The second index measured in the $f_\mathrm{diff}$ spectrum, that roughly describes the shape and location of an emission feature, is the asymmetry index defined as:
\begin{equation}
	\label{equ:hab_asym}
	Asymmetry = \frac{|EW_\mathrm{red}| - |EW_\mathrm{blue}|}{|EW_\mathrm{red}| + |EW_\mathrm{blue}|},
\end{equation}
where $|EW_x|$ represents the equivalent width of the absolute difference $|f_\mathrm{diff}|$ on the red and blue side of the central wavelength of the investigated Balmer line. By this definition, a line that is, as a whole, moved to the redward side would have this index equal to $1$, whilst if it was moved to the blueward side, the index would instead equal $-1$. The distribution of the asymmetry index values for the most prominent and unflagged (see Section \ref{sec:flagging}) emitters is shown in Figure \ref{fig:hab_asym}, where a strong correlation between the asymmetry of H$\alpha$ and H$\beta$ lines is evident. As the H$\beta$ line in most cases produces a much weaker or even no emission feature, its asymmetry is much harder to measure. That is evident in Figure \ref{fig:hab_asym} where its index is scattered a around $0$, except for the most asymmetric cases. The distribution of the H$\alpha$ asymmetry is much more uniform outside the central symmetric region. From this index, we can roughly classify the source of the emitting component as a chromospheric origin would produce a centered component with an asymmetry index close to $0$. Everything outside the central region in Figure \ref{fig:hab_asym}, defined by the circle with a radius of $0.25$, could be thought to be of an extra-stellar origin as lines are not perfectly aligned. The used thresholding radius value of $0.25$ was defined by observing Figure \ref{fig:hab_asym} to encircle the main over-density of almost symmetric emission profiles.

% Like with figures 6 and 7, do we have expectations for how the data should look in figures 8 and 9? 
\begin{figure}
	\centering
	\includegraphics[width=\columnwidth]{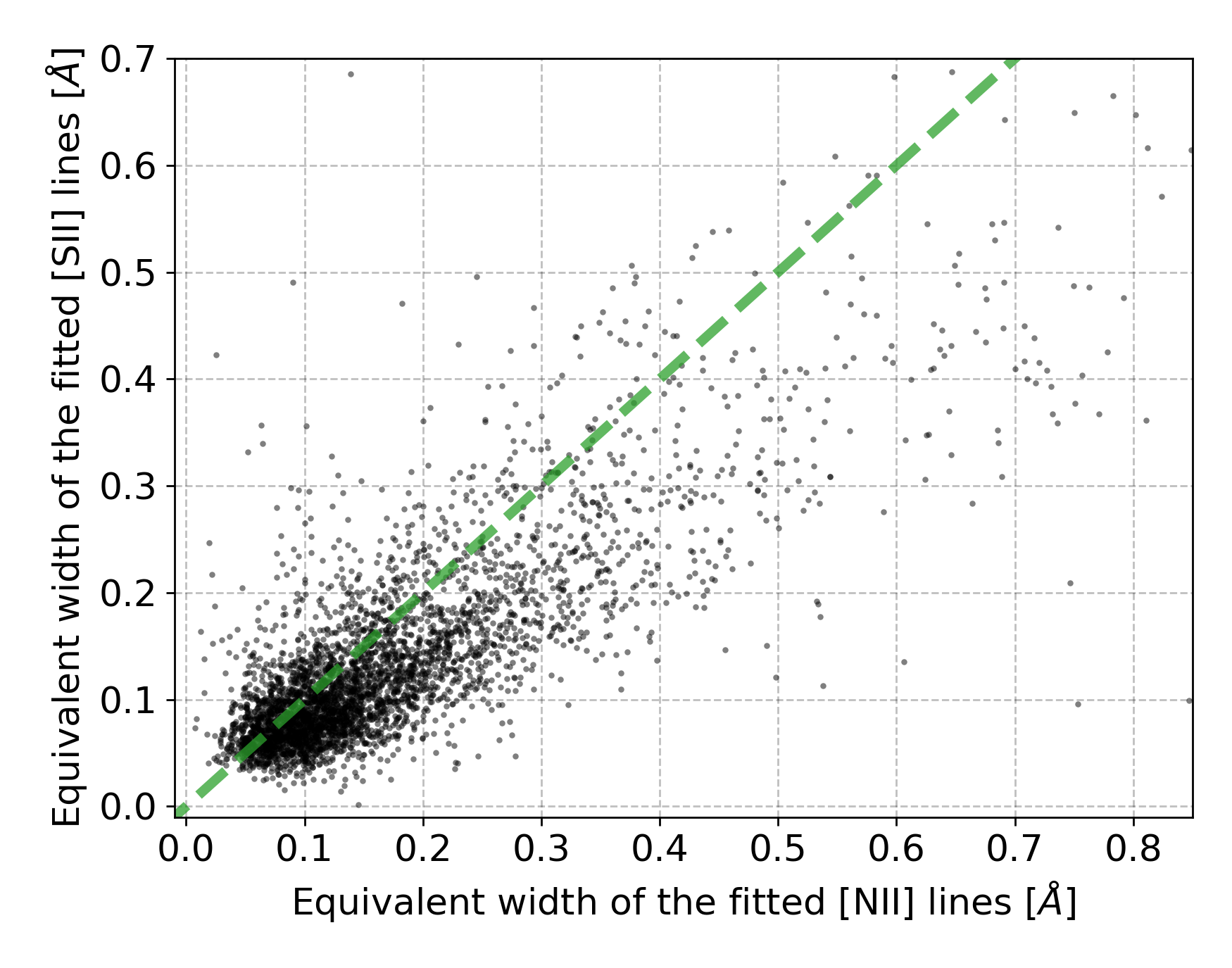}
	\caption{Correlation between the strengths of the nebular contributions from both elements. Shown are only cases with a small difference in the determined radial velocities as shown in Figure \ref{fig:sii_nii_rv}.}
	\label{fig:sii_nii_ew}
\end{figure}

\subsection{Detection of nebular contributions}
\label{sec:nebularemis}
Due to the multiple possible origins of H emission lines \citep{2007ASSL..342.....K}, we also attempted to detect the extra-stellar nebular contributions of nearby rarefied gas. Its presence is expressed as forbidden emission lines in addition to the H emission. The spectral coverage  of the HERMES red arm enables us to observe doublets of [SII] ($6548.03$ and $6583.41$ \AA), and [NII] ($6716.47$ and $6730.85$ \AA). Having usually a weak emission contribution that could possibly be blended with nearby absorption lines, they are most easily detected when we remove the expected reference spectrum from the observed one (resulting in $f_\mathrm{diff}$). To automatically detect the emission strength and position of both doublets, we independently fitted two Gaussian functions with the same radial velocity shift for each element to $f_\mathrm{diff}$. Because the contributing medium is not necessarily physically related to the observed object, its radial velocity could be different, therefore it was treated as a free parameter in our fit. Two independent velocities, one for each of the two doublets, give us an indication of a spurious or unreliable fit component if their difference is large. To filter out outliers, we adopted a threshold of $15$~\kms\ on their velocity difference. Some of the discarded outliers might be correct detections because few of the spectra show two or more peaks for each nebular line which might point to a contribution of multiple clouds with different radial velocities. Such cases are not fully accounted for by the fitting algorithm that only identifies the strongest emission.

\begin{figure}
	\centering
	\includegraphics[width=\columnwidth]{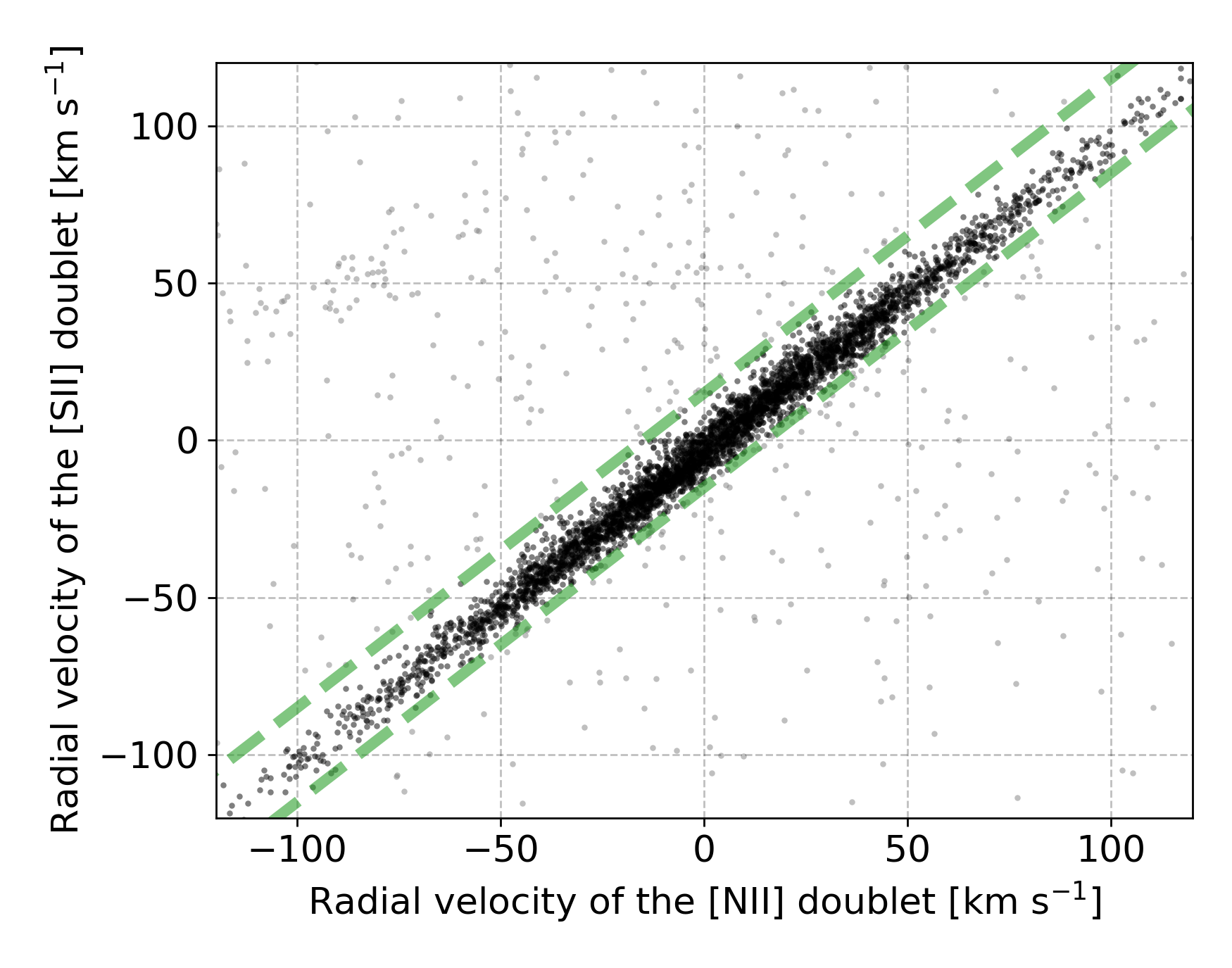}
	\caption{Correlation between radial velocity of both assessed nebular contributions that are observable in the red arm of the HERMES spectrum. Shown are only cases with at least three detected forbidden lines. The grey dots were further discarded as their absolute difference between velocities is more than $15$~\kms. The limiting thresholds are visualized by dashed linear lines. Plotted velocities are measured in the stellar rest frame and therefore grouped towards zero velocity, meaning they are moving together with the star.}
	\label{fig:sii_nii_rv}
\end{figure}

In the absence of additional fitting constraints, the routine might also find two noise peaks and lock onto them. Therefore, we put an arbitrarily selected detection threshold ($0.05$ of relative flux) on a minimum amplitude of the fitted forbidden lines to be counted as detected. The result from this fitting and analysis procedure is a number of successfully detected peaks per element and their combined equivalent widths (EW([NII]) and EW([SII])), reported in the final published table (see Table \ref{tab:results}). To filter out some possible miss-detection, we count a spectrum as having nebular lines when at least three nebular lines above the threshold were detected. The correlations for measured radial velocities and equivalent widths of identified objects with nebular emission are given in Figure \ref{fig:sii_nii_ew} and \ref{fig:sii_nii_rv} respectively.

The radial velocities of both doublets shown in Figure \ref{fig:sii_nii_rv} give us a first impression that the gas dynamics of the elements in all observed clouds is nearly coincident, but elements are moving at slightly different velocities. This velocity offset, but in the opposite direction, was also observed by \citet{2016A&A...591A..74D, 2017A&A...604A.135D} who attributed it to the uncertainties in their adopted line wavelengths, that are slightly different to ours (less than $0.05$~\AA), causing the velocity points to be located either above or under the identity line in Figure \ref{fig:sii_nii_rv}. Additionally, the plot reveals that the majority of the gas clouds have a different radial velocity than stars behind or inside a cloud.

As we are working with fully reduced normalised spectra, with inclusion of sky background removal, the detection procedure would, in the case of an ideal background removal, not detect emission due to nebular clouds. As the measured flux of the nebular contribution is very unlikely the same for object and because of the physical separation of the sky fibres (see next Section \ref{sec:skyemis} and \citet{2017MNRAS.464.1259K}), the ideal cases are very rare. Similarly, the densities and the temperatures of such nebular clouds, extracted from corrected spectra could be influenced by the extraction pipeline and were therefore not performed in our case. 

\begin{figure}
	\centering
	\includegraphics[width=\columnwidth]{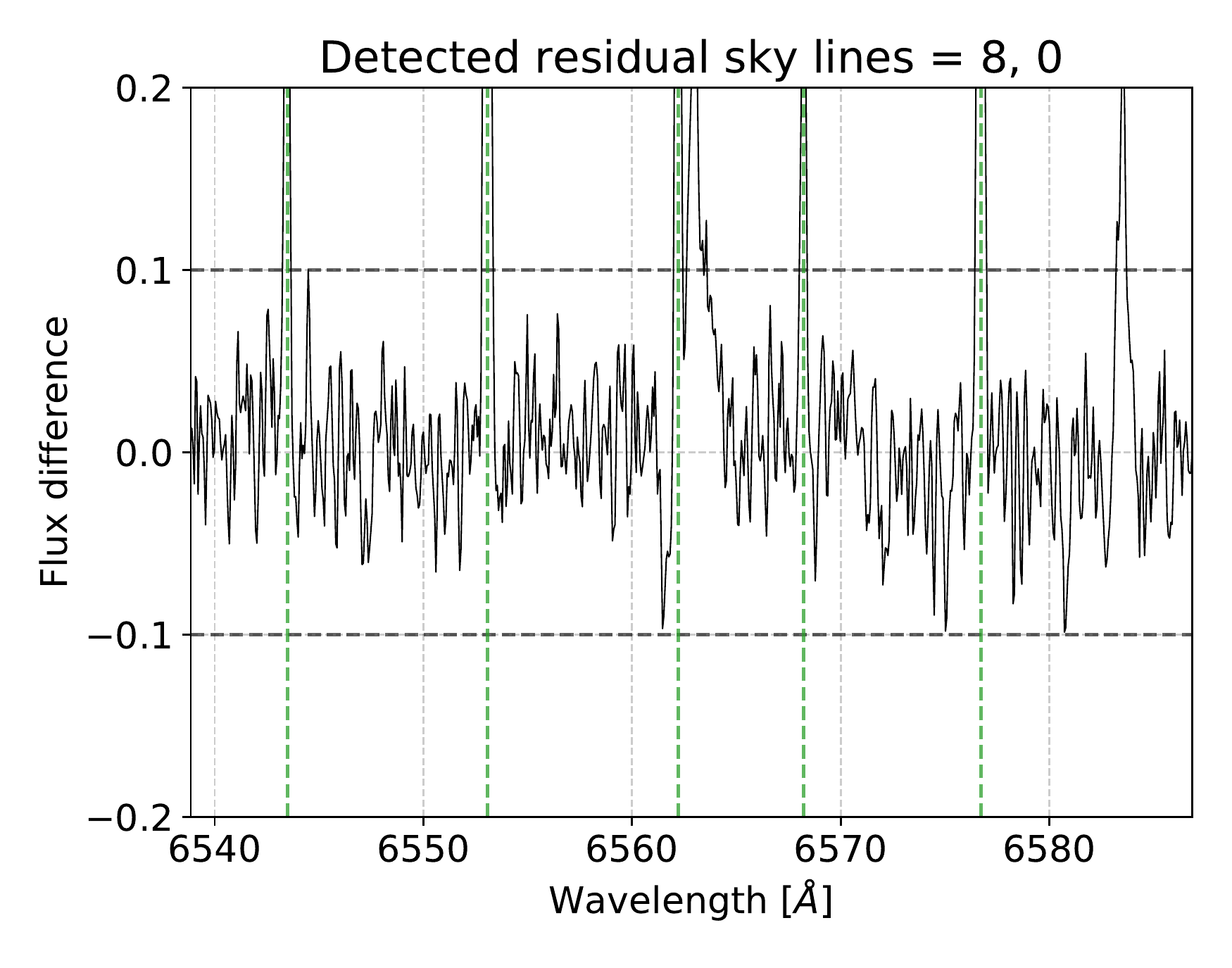}
	\caption{Sky emission lines are most evident after spectra subtraction in $f_\mathrm{diff}$. Green vertical dashed lines represent expected locations of emission lines in the rest frame of an observed star. The middle sky line in this plot falls inside the actual H$\alpha$ emission feature and changes its shape from single- to double-peaked and consequently modifies the measured equivalent width. Upper and lower thresholds for detection are given by the bold horizontal dashed lines. The number of detected under- and over-corrected sky lines in this order is given above the plot.}
	\label{fig:skyemiss}
\end{figure}

The strength of the identified lines, measured by their equivalent widths, is shown in Figure \ref{fig:sii_nii_ew}. This shows high a degree of correlation, where on average [SII] lines have lower strength than [NII] lines.

\subsection{Identification of sky emission lines}
\label{sec:skyemis}
Attributing a limited and relatively low number of the HERMES fibres to monitor the sky in hopefully star and galaxy free regions, imposes limitations to a quality of the sky background removal in the GALAH reduction pipeline \citep{2017MNRAS.464.1259K}. As the sky spectrum is sampled at $25$ distinct locations over the whole $2^\circ$ diameter field, it must be interpolated for all other fibre locations that are pointing towards stellar sources. Depending on the temporal and spatial variability of weather conditions, and possible nebular contributions, interpolation may produce an incorrect sky spectrum that is thereafter removed from the observed stellar spectra.

In most cases, this does not influence the spectral analysis, unless one of the strongest sky emission lines falls in a range of the analysed stellar line. For us, the most problematic sky emission line, which can alter the shape of the H$\alpha$ profile, is located at $6562.7598$~\AA\ \citep[our list of sky emission lines was taken from][]{2003A&A...407.1157H}. As it can get blended with a real emission feature of the H$\alpha$ or simulate its presence, we try to estimate the impact of the sky residual in the spectrum from multiple nearby emission lines. First, we select only the strongest sky emitters \citep[with parameter \texttt{Flux}~$\ge$~$0.9$ in][]{2003A&A...407.1157H} and shift their reference wavelength into a stellar rest frame. After that, we use a simple thresholding (see Figure \ref{fig:skyemiss}) to estimate their number. By the thresholding procedure, we want to simultaneously catch over- and under-corrected stellar spectra.

When a sufficient number ($\ge$~$4$) of strong residual sky lines with a normalised flux above $10$\% is detected, a quality flag (see Section \ref{sec:flagging}) is raised, warning a user that the equivalent width of the H$\alpha$ emission could be affected by uncorrected sky emission. As this potential contamination is present only in the red HERMES arm, we do not check for spurious strong emitters in the region around the H$\beta$ line.

\begin{figure}
	\centering
	\includegraphics[width=\columnwidth]{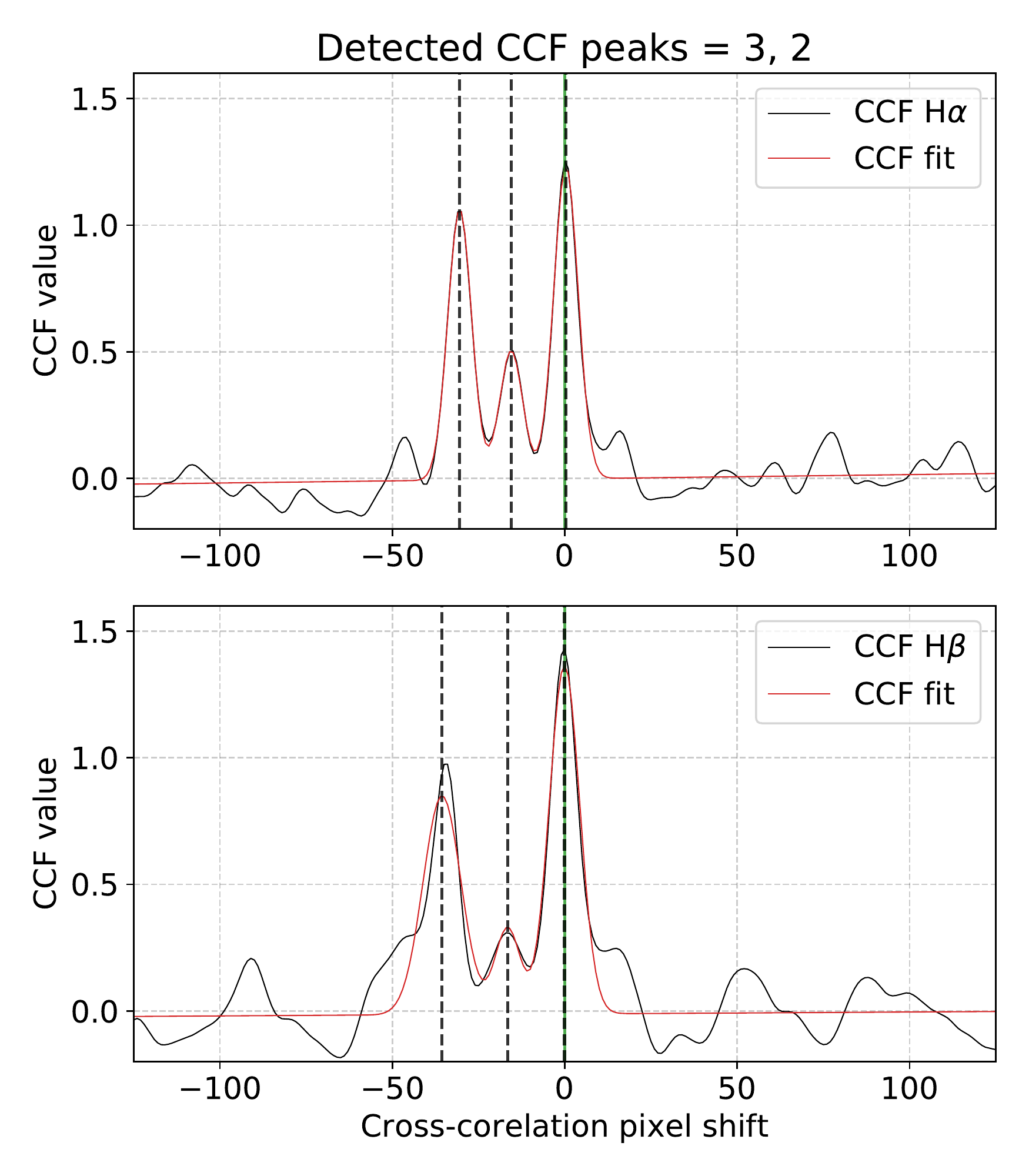}
	\caption{Detection of a spectroscopic binary candidate by cross-correlating observed spectrum with its reference spectrum. Three Gaussian functions that are fitted to the resulting CCF (black solid curve) are depicted by their means (dashed vertical lines) and their best fitting sum in red. Presented are CCFs for the red arm in the top panel and for the blue arm in the bottom panel. Number of detected peaks for both arms is given above the figure.}
	\label{fig:sb2ccf}
\end{figure}

\begin{table}
	\caption{Quality binary flags produced during different steps of our detection and analysis pipeline. Lower value of the flag represents lower significance to the quality of detection and classification. The final reported \texttt{flag} value in Table \ref{tab:results} is a sum of all raised binary quality flags.}
	\label{tab:flags}
	\begin{tabular}{r l}
		\hline
		Flag & Description \\
		\hline
		128 & Reference spectrum for the H$\alpha$ range does not exist.\\
		64 & Reference spectrum for the H$\beta$ range does not exist.\\
		32 & Large difference between reference and observed\\ 
		& spectrum in the red arm of a spectrum. Median\\
		& squared error (MSE) between them was $\ge 0.002$\\
		16 & Large difference between reference and observed\\
		& spectrum in the blue arm of a spectrum.\\
		& MSE was $\ge 0.008$.\\
		8 & The spectrum most likely contains duplicated\\
		& spectral absorption lines of a resolved SB2 binary\\
		& Binarity was detected in both arms.\\
		4 & Possible strong contamination by sky emission\\
		& features. 4 or more residual sky lines were detected.\\
		& Could be a result of under- or over-correction.\\
		2 & Wavelength solution (or determined radial velocity)\\
		& might be wrong in the red arm of the spectrum.\\
		& Determined from cross-correlation peak between\\
		& observed and reference spectra.\\
		1 & Wavelength solution (or determined radial velocity)\\
		& might be wrong in the blue arm of the spectrum.\\
		\hline
	\end{tabular}
\end{table}

\subsection{Determination of spectral binarity}
During the inspection of our initial results, we noticed that the spectra of spectroscopically resolved binary stars (SB2) produce a mismatch between observed and reference spectra whose $f_\mathrm{diff}$ have a profile similar to the P Cygni or inverted P Cygni profile \citep{1979ApJS...39..481C} that is often observed in emission-line objects. To detect SB2 candidates, we performed cross-correlation between the reference and the observed spectra, disregarding the wavelength range of $\pm10$\AA\ around the centre of the Balmer lines to avoid broadening of the cross-correlation function (CCF) peak. Cross-correlation was performed independently for both (the blue and red) HERMES spectral arms. The resulting CCF, shown as the black curve in Figure \ref{fig:sb2ccf}, was fitted by three Gaussian functions, centred at three strongest peaks, to describe its shape. The location, amplitude and width of those peaks were assessed to determine the number of stellar components in the spectrum. When fitting three peaks, there is a possibility of finding triple stars and distinguishing them from binaries. Every spectral arm with more than one prominent peak was marked as potential SB2 detection in the final results (see Table \ref{tab:results}), where binarity indication is given independently for both arms. Nevertheless, the results of the blue arm (column \texttt{SB2\_c1}) are more trustworthy because of the higher number of absorption lines in the red arm (column \texttt{SB2\_c3}). For even greater completeness of detected SB2 candidates, a list of analyzed binaries, compiled by \cite{2020arXiv200500014T} can be used. They combined unsupervised spectral dimensionality reduction algorithm t-SNE and semi-supervised CCF analysis \citep{2017A&A...608A..95M} to compile their list of SB2 binaries. After their analysis, they discarded spectra that were falsely identified as SB2 by their detection procedures.

\begin{figure*}
	\centering
	\includegraphics[width=\textwidth]{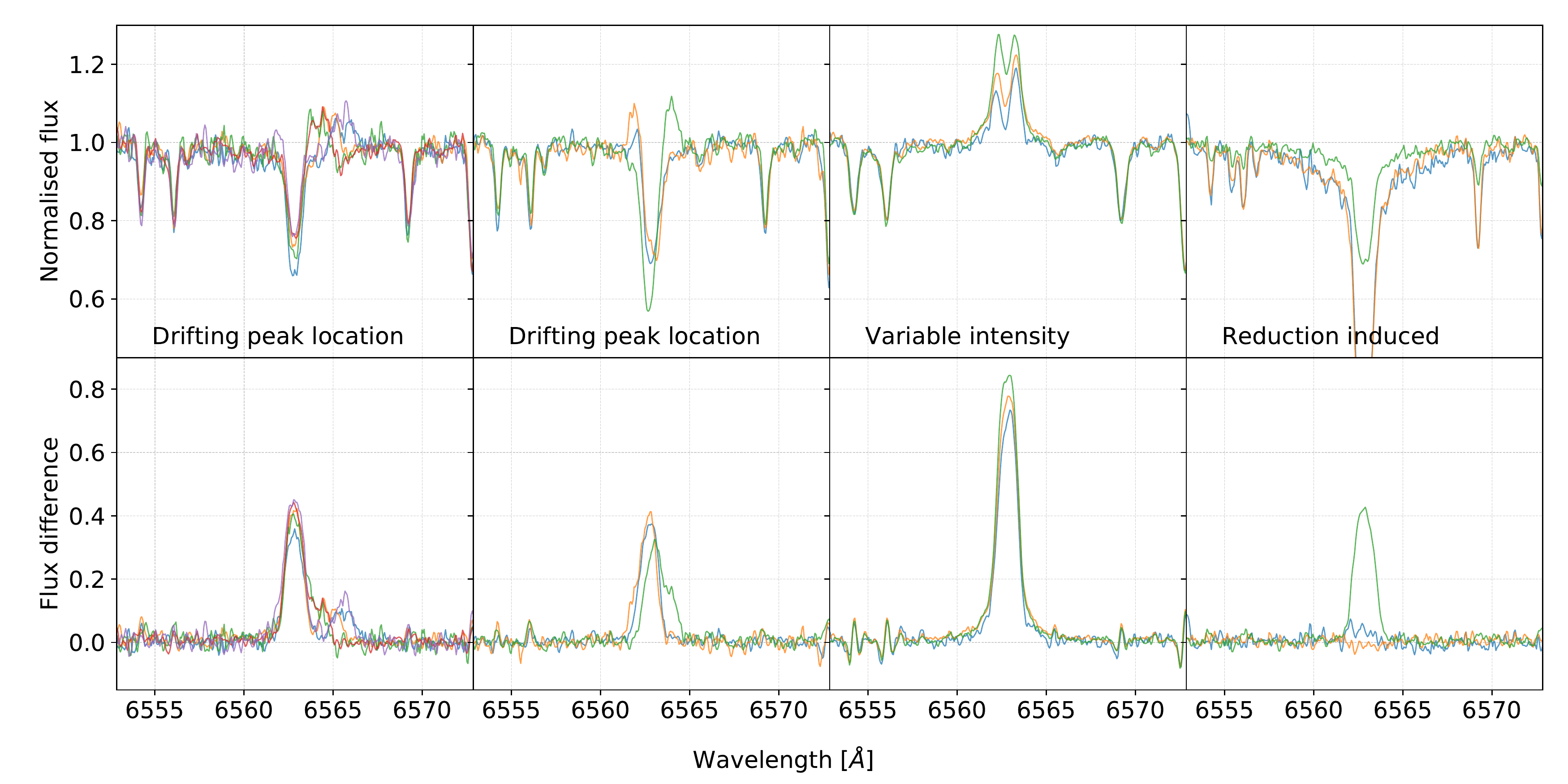}
	\caption{A sample of objects with repeated observations, where at least one of the normalised spectra (top row) contains strong H$\alpha$ emission detected by comparison towards reference spectrum (bottom row). The first two objects (or columns) show shifting location of an additional emission component peak, and the last two varying degree of its strength. The last example is most likely a result of a miss-reduction as not only H$\alpha$, but also other absorption lines show reduced strength. The existence of this problem is confirmed by other objects in the same field as majority of them show the same tendency of having weaker absorption lines across the spectrum.}
	\label{fig:temporal_variab}
\end{figure*}

An unexpected result of this binarity search was the realization that some reduced spectra show duplicated lines only in the red arm or even stranger, only in a smaller subsection of it. After a thorough investigation, we uncovered that this effect is caused by improper treatment of fibre cross-talk while extracting spectra from the original 2D image \citep{2017MNRAS.464.1259K}. A partial culprit of this is also a poorer focus in the red arm. Therefore if only flag \texttt{SB2\_c3} is set, and not \texttt{SB2\_c1}, this can be used as an indication of the above reduction effect.

Additionally, the highest peak of our CCF function is used to determine the correctness of the wavelength calibration during the reduction of the spectra \citep{2017MNRAS.464.1259K}. If the peak is shifted by more than five correlation steps (equalts to about 13~\kms) from the rest wavelength of the reference spectrum, the quality flag (see Section \ref{sec:flagging}) is raised, warning the user that the derived radial velocity, equivalent width, and asymmetry index might be wrong in the respective arm as both spectra were not aligned ideally.

\subsection{Resulting table}
\label{sec:results}
The emission indices and other computed parameters are collected in Table \ref{tab:results}. The complete table is available in electronic form at the CDS. An excerpt of the published results, containing a subset of 30 rows and 11 most interesting columns for the strongest unflagged emitters is given in Table \ref{tab:results_values}.

As we do not perform any quality cuts on our results, a suggested set of limiting parameter thresholds and quality flags is provided in Section \ref{sec:flagging}. Their use depends on user specific requirements and the analysed science case.

\subsection{Flagging, quality control and results selection}
\label{sec:flagging}
The above described pipeline runs blindly on every successfully reduced spectrum \citep[\texttt{guess\_flag} = 0, for details see][]{2017MNRAS.464.1259K}, and could therefore produce wrong or misleading results for some spectra. To have the ability to filter out such possible occurrences, we created a set of warning flags for different pipeline steps that are listed and described in detail in Table \ref{tab:flags}. An interested user can base their selection of results according to the desired confidence level and a physical question of interest. The cleanest set of $10,364$ H$\alpha$ emission stars can be produced by selecting unflagged stars that do not show any signs of possible binarity, defined such that parameter \texttt{emiss} in the published Table \ref{tab:results} is set to one (the equivalent of true). To be included among the cleanest set of detections, we considered only spectra whose \texttt{Ha\_EW}~>~$0.25$~\AA. Below this limit, we are less confident in marking an object as having an emission feature because visual inspection showed that this strength could be mimicked by spectral noise, the uncertainty of the reference spectrum, or induced by the reduction pipeline. This selection criteria at the same time discards the weakest chromospheric components, which might be of great interest for specific studies. If the user is interested only in stronger emitters, the threshold should be raised to \texttt{Ha\_EW}~>~$0.5$~\AA\ or above.

\begin{figure*}
	\centering
	\includegraphics[width=\textwidth]{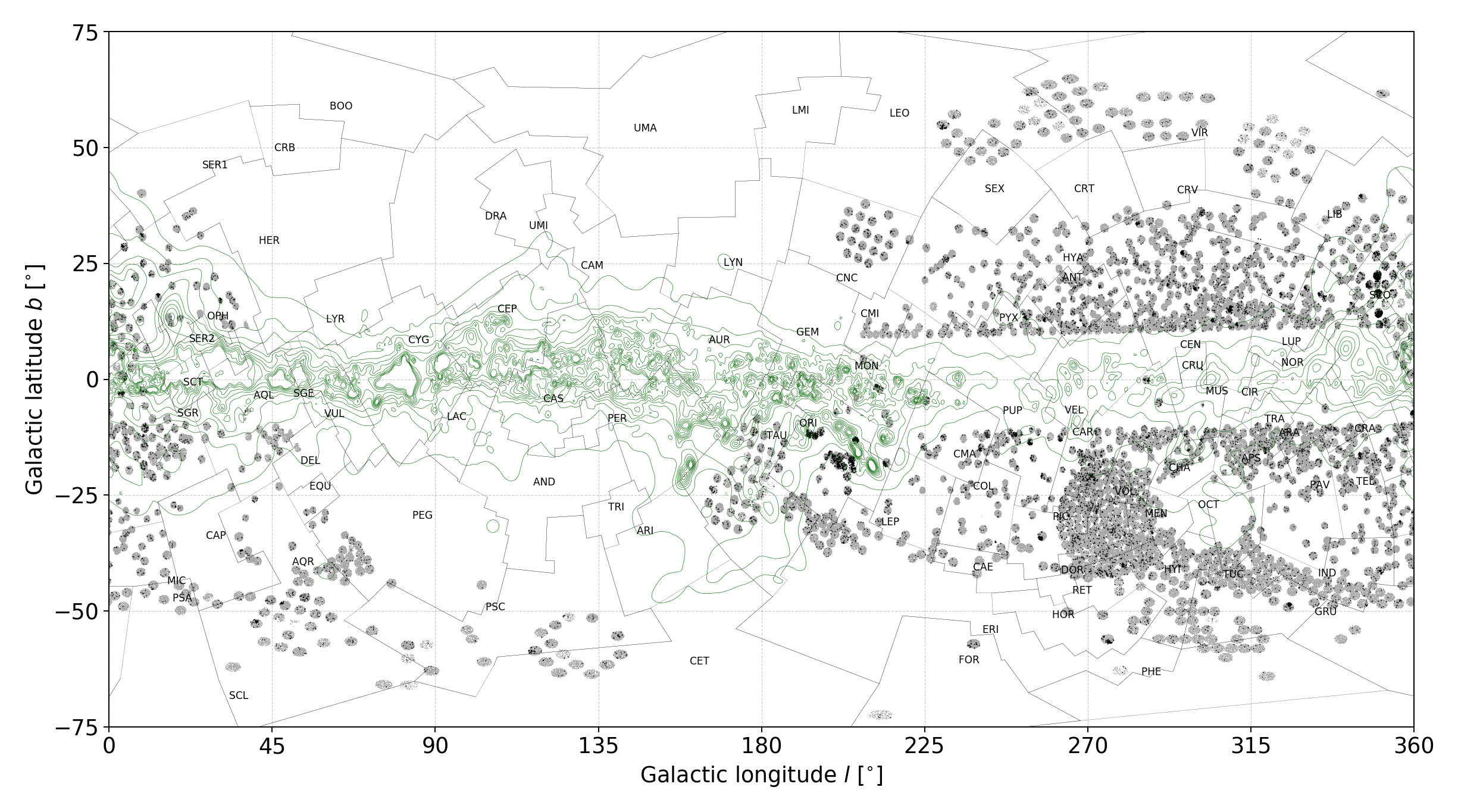}
	\caption{Spatial distribution of stars with detected emission Balmer emission profiles. Grey areas represent regions that were observed and analysed in this paper. The green lines represent location of equal reddening in steps of $0.1$ magnitude at the distance of $2$~kpc. Reddening data were taken from results published by \citet{2017A&A...606A..65C}. For readability, no isoline is shown above the reddening of 1~magnitude. Constellation boundaries were taken from \citet{const_data}. Locations of their designations are defined by median values of constellation polygon vertices.}
	\label{fig:spatialemission}
\end{figure*}

The published Table \ref{tab:results} also contains a flag that describes whether the spectrum is considered to contain an additional nebular contribution. Such spectra can be filtered out by choosing the parameter \texttt{nebular} to be equal to $1$. To compile this less restrictive list of $4004$ spectra, we selected entries with at least three prominent forbidden emission lines (\texttt{NII}~$+$~\texttt{SII}~$\ge$~$3$) and a small difference in their measured radial velocities ($|$\texttt{rv\_NII}~$-$~\texttt{rv\_SII}$|$~$\le$~$15$~\kms).

\section{Temporal variability}
\label{sec:temporal}
The strategy of the GALAH survey is to observe as many objects as possible, and as a result, not many repeated observations were made. The repeated fields were mostly observed to assess the stability of the instrument. Time spans between observations are therefor on the orders of days or years. This limits the possibility of finding a variable object greatly, but still enables us to discover potential interesting objects and diagnose analysis issues.

\begin{figure}
	\centering
	\includegraphics[width=\columnwidth]{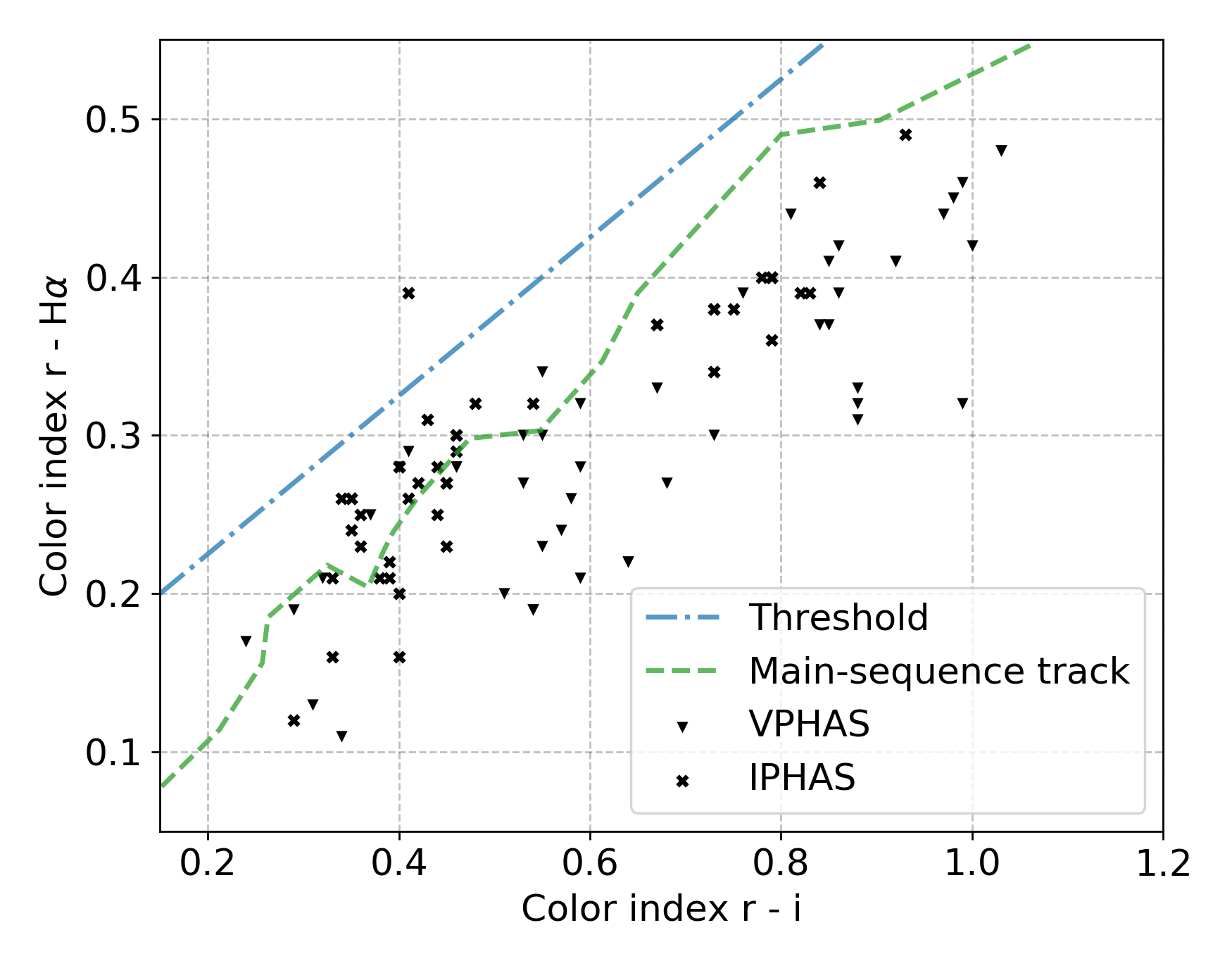}
	\caption{r - H$\alpha$ versus r - i color-color diagram using combined IPHAS and VPHAS photometry for detected emission stars. The dashed green line represents the unreddened main-sequence track tabulated by \citet{2014MNRAS.440.2036D}. An empirical threshold, shown by dash-dotted blue line, can be used to distinguish non-accreting and accreting objects above the line \citep{2018A&A...609A..10V}.}
	\label{fig:iphas_vphas}
\end{figure}

To find possible emission stars with repeated observations, we selected stars with repeats, among which at least one spectrum was identified to harbour a stronger (\texttt{Ha\_EW}~>~$0.5$~\AA) unflagged emission feature. This selection produced $621$ stars, having between $2$ and $9$ observations. To be confident about the observed variability, we visually inspected the observed and the reference spectra of $208$ stars with at least three observations. A subset of these spectra are shown in Figure \ref{fig:temporal_variab}, where we present typical types of variability discovered by visual inspection. The types can roughly be described as shape transformation (e.g. change from single- to double-peak or P Cygni emission profile), peak location shift, intensity change, and possible reduction issue.

In the sample of 208 stars, whose spectra were visually inspected, we found that $\sim20$\% of the inspected spectra display a stable H$\alpha$ profile. Noticeable profile shape transformation was observed in $\sim10$\% of the cases, and peak location change in $\sim5$\% of the cases. Some degree of emission intensity change was noticed for $\sim40$\% of the cases. Visually similar is reduction induced variability (see the rightmost panel in Figure \ref{fig:temporal_variab}), observed for $\sim25$\% of all inspected repeated observations. In the case of multiple observations of the same star, we can distinguish between the last two profile changes (intrinsic and reduction induced intensity change) by looking at the whole spectrum to inspect whether variability is also exhibited in other absorption lines as shown by the last example in Figure \ref{fig:temporal_variab}. That kind of reduction induced variability is limited to a few observed fields. 

\section{Discussion and conclusions}
\label{sec:discussion} 
In this paper, we describe the development and application of a neural network autoencoder structure that is able to extract the most relevant latent features from the spectrum. Low feature dimensionality contains only the most basic spectral informations that are used to reconstruct a non-peculiar spectrum with the same physical parameters as the input spectrum.

Our method of differential spectroscopy is one of the most widely used approaches to find peculiar spectral features that are not found in normal stars. As a part of this paper, we showed that a dense autoencoder neural network structure can be reliably used for generation of non-peculiar reference spectra if trained on a large set of normal spectra. With the additional exclusion of our detected emission-line stars, the training set could iteratively be further cleaned of peculiar stars before training the network. As all the information about the spectral look is contained in the real flux values, there is no need to add additional convolutional layers for the extraction of more complex spectral shapes. 

By identifying significant residuals after subtracting the generated reference spectra from the observed spectra, we detected emission star candidates in the GALAH fields all over the sky. Figure \ref{fig:spatialemission} shows that we can identify few locations with a higher density of detected emission-line objects. The position of emission-line objects coincides with regions of young stars such as the Orion complex, Blanco 1, Pleiades, and other possibly random over-densities of interstellar gas and dust. Detected nebular emission in stellar spectra, shown in Figure \ref{fig:spatialnebular}, coincide with large visually-identified nebular clouds \citep[by comparing detected locations with the red all-sky photographic composite of The Second Digitized Sky Survey, described by][]{2000ASPC..216..145M} such as the Antares Emission nebulae, clouds around $\pi$ Sco and $\delta$ Sco, Barnard's loop, Carina Nebula, nebulae around $\lambda$ Ori, nebular veils in the constellations of Puppis, Pyxis and Antlia, and other less prominent features.

By combining our detections with additional auxiliary data sets, we can start exploring more detailed physical explanations of the observed emissions and their structure. Among them are two specific photometric surveys, VPHAS \citep{2014MNRAS.440.2036D} and IPHAS \citep{2008MNRAS.384.1277W} which were designed to detect and study emission-line sources close to the Galactic plane. Because of their positional selection function, their combined photometric data are available only for $4431$ GALAH spectra. Of these, the spectroscopically confirmed emission stars are shown in Figure \ref{fig:iphas_vphas}, whose color-color diagram can be used to infer accreting objects.

Our detected emission spectra have a broad range of emission components - these range from very strong to barely detectable chromospheric emission component whose identification can be mimicked or masked at multiple steps of the analysis and data preparation. To limit the number of false-positive classifications due to reduction and analysis limitations, we focused on stronger components (\texttt{Ha\_EW}~>~$0.25$~\AA) whose existence can be confirmed visually. Because that kind of process would be slow for the whole sample, we introduced quality flags that can be used to filter out unwanted or specific cases. Additionally, the stability of the spectra and emission features was investigated by repeated observations of the same objects. Among them, we observed different variability types, of which one could be attributed to the data reduction pipeline, limiting the confidence of finding weak emission profiles in the spectra.

To reliably detect even the weakest chromospheric emissions, uncertainty of the used reference spectra must be well known as well. By showing that the proposed neural network structure can be used as intended, we are looking into possibilities to improve our methodology using variational autoencoder. Its advantage lies in the possibility of simultaneous determination of a reference spectrum and its uncertainty which would enable uncertainty estimation of the measured emission-line indices.

%%%%%%%%%%%%%%%%%%%%%%%%%%%%%%%%%%%%%%%%%%%%%%%%%%

\section*{Acknowledgements}
This work is based on data acquired through the Australian Astronomical Observatory, under programmes: A/2014A/25, A/2015A/19, A2017A/18 (The GALAH survey); A/2015A/03, A/2015B/19, A/2016A/22, A/2016B/12, A/2017A/14 (The K2-HERMES K2-follow-up program); A/2016B/10 (The HERMES-TESS program); A/2015B/01 (Accurate physical parameters of Kepler K2 planet search targets); S/2015A/012 (Planets in clusters with K2). We acknowledge the traditional owners of the land on which the AAT stands, the Gamilaraay people, and pay our respects to elders past and present.

The Digitized Sky Surveys were produced at the Space Telescope Science Institute under U.S. Government grant NAG W-2166. The images of these surveys are based on photographic data obtained using the Oschin Schmidt Telescope on Palomar Mountain and the UK Schmidt Telescope. The plates were processed into the present compressed digital form with the permission of these institutions.

K\v{C}, TZ, and JK acknowledge financial support of the Slovenian Research Agency (research core funding No. P1-0188 and project N1-0040) and European Space Agency (PRODEX Experiment Arrangement No. C4000127986).

JBH is funded by an ARC Laureate Fellowship. 

SB acknowledges funds from the Australian Research Council (grants DP150100250 and DP160103747) as well as from the Alexander von Humboldt Foundation in the framework of the Sofja Kovalevskaja Award endowed by the Federal Ministry of Education and Research. Parts of this research were supported by the Australian Research Council (ARC) Centre of Excellence for All Sky Astrophysics in 3 Dimensions (ASTRO 3D), through project number CE170100013. 

%%%%%%%%%%%%%%%%%%%%%%%%%%%%%%%%%%%%%%%%%%%%%%%%%%



\bibliographystyle{mnras}
\bibliography{bibliography}

%%%%%%%%%%%%%%%%%%%%%%%%%%%%%%%%%%%%%%%%%%%%%%%%%%

%%%%%%%%%%%%%%%%% APPENDICES %%%%%%%%%%%%%%%%%%%%%

\appendix

\section{Table description}
In the Table \ref{tab:results} we provide a list of metadata available for every object analysed using the methodology described in this paper. The complete table of detected objects and its metadata is available only in electronic form at the CDS and at publishers website.

\begin{table*}
	\caption{List and description of the fields in the published catalogue of analysed objects.}
	\label{tab:results}
	\begin{tabular}{l c l}
		\hline
		Column & Unit & Description \\
		\hline
		\texttt{source\_id} & & \G\ DR2 star identifier \\
		\texttt{sobject\_id} & & GALAH internal per-spectrum unique id \\
		\texttt{ra} & deg & Right ascension coordinate from 2MASS\\
		\texttt{dec} & deg & Declination coordinate from 2MASS\\
		\texttt{Ha\_EW} & \AA & Equivalent width of a difference between observed and template spectrum in the range \\
		& & of $\pm3.5$~\AA\ around the H$\alpha$ line \\
		\texttt{Hb\_EW} & \AA & Same as the \texttt{Ha\_EW}, but for the H$\beta$ line \\
		\texttt{Ha\_EW\_abs} & \AA & Equivalent width of an absolute difference between observed and template spectrum in \\
		& & the range of $\pm3.5$~\AA\ around the H$\alpha$ line \\
		\texttt{Hb\_EW\_abs} & \AA & Same as the \texttt{Ha\_EW\_abs}, but for the H$\beta$ line\\
		\texttt{Ha\_W10} & \kms & Width (in \kms) of the H$\alpha$ emission feature at $10$\% of its peak flux amplitude \\
		\texttt{Ha\_EW\_asym} & & Value of asymmetry index for the H$\alpha$ line \\
		\texttt{Hb\_EW\_asym} & & Value of asymmetry index for the H$\beta$ line \\
		\texttt{SB2\_c3} & & Was binarity detected in the red arm \\
		\texttt{SB2\_c1} & & Was binarity detected in the blue arm \\
		\texttt{NII} & & Number of detected [NII] peaks in the doublet \\
		\texttt{SII} & & Number of detected [SII] peaks in the doublet \\
		\texttt{NII\_EW} & \AA & Equivalent width of a fitted Gaussian profiles to the [NII] emission features \\
		\texttt{SII\_EW} & \AA & Same as the \texttt{NII\_EW}, but for the [SII] doublet \\
		\texttt{rv\_NII} & \kms & Intrinsic radial velocity of the [NII] doublet, corrected for the barycentric and stellar velocity \\
		\texttt{rv\_SII} & \kms & Same as \texttt{rv\_NII}, but for the [SII] doublet \\
		\texttt{nebular} & & Is spectrum considered to have an additional nebular component \\
		\texttt{emiss} & & Is spectrum considered to have an additional H$\alpha$ emission component \\
		\texttt{flag} & & Sum of all bitwise flags raised for a spectrum \\
		\hline
	\end{tabular}
\end{table*}

\section{Additional figures}
In order to increase the readability and transparency of the text, additional and repeated plots are supplied as appendices to the main text. 

\begin{figure*}
	\centering
	\includegraphics[width=\textwidth]{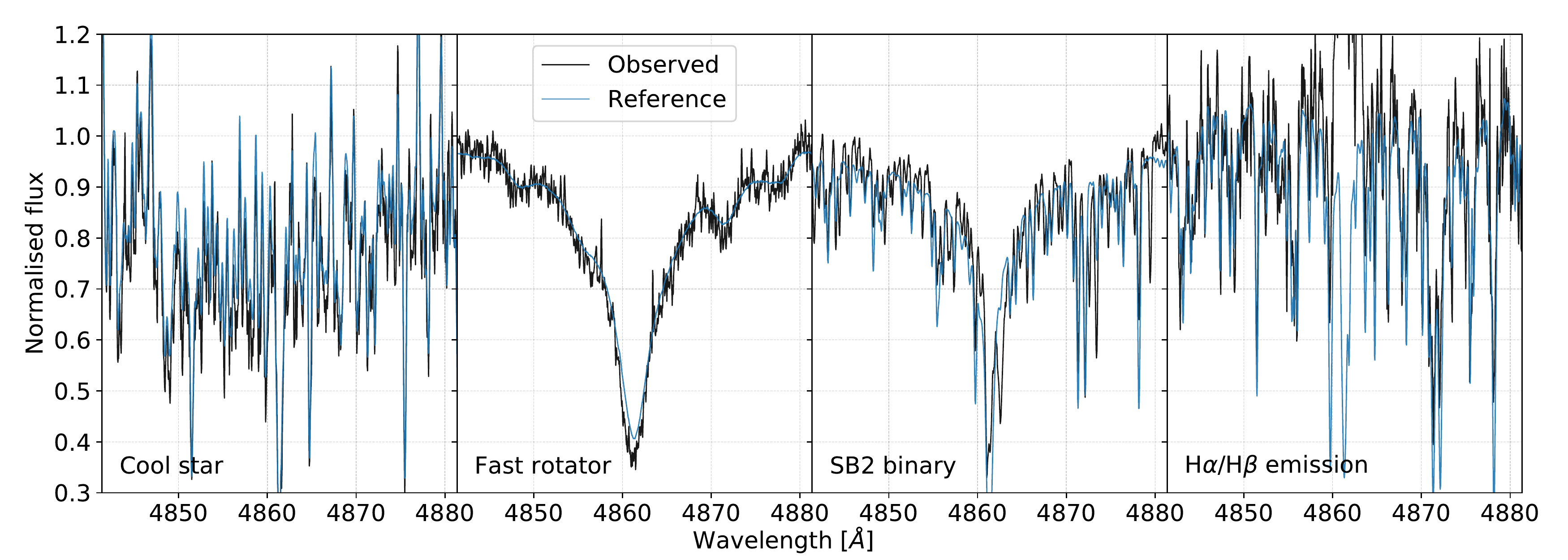}
	\caption{Same plots and objects as in Figure \ref{fig:refann} but for the blue spectral arm.}
	\label{fig:refann2}
\end{figure*}

\begin{table*}
	\caption{Excerpt of 30 strongest unflagged emitters from the published table presented in detail by Table \ref{tab:results}. The rest of the table can be downloaded in electronic form CDS service and publishers' website.}
	\label{tab:results_values}
	\begin{tabular}{c | c | c | c | c | c | c | c | c | c | c}
\hline
source\_id & Ha\_EW & Ha\_EW\_abs & Ha\_W10 & Ha\_EW\_asym & NII & SII & NII\_EW & rv\_NII & rv\_SII & flag \\
\hline
3337923100687567872 & 5.37 & 5.37 & 373.63 & 0.36 & 1 & 0 & 0.05 & -30.22 & 23.54 & 0 \\
3217769470732793856 & 5.06 & 5.06 & 252.48 & 0.07 & 0 & 1 & 0.01 & -11.32 & 35.71 & 0 \\
4660266122976778240 & 4.51 & 4.51 & 199.66 & 0.15 & 0 & 0 & 0.02 & -310.50 & -231.18 & 0 \\
3340892714091577856 & 4.35 & 4.35 & 205.94 & -0.06 & 2 & 2 & 0.20 & -24.73 & -27.14 & 0 \\
3336365097008009216 & 4.14 & 4.14 & 188.62 & -0.08 & 0 & 0 & 0.07 & -76.99 & -43.80 & 0 \\
3217804483306125824 & 4.02 & 4.02 & 152.24 & -0.03 & 0 & 1 & 0.01 & -85.06 & 28.70 & 0 \\
3214742618300312064 & 3.97 & 3.98 & 277.85 & -0.08 & 0 & 0 & 0.01 & -63.06 & -32.82 & 0 \\
6243142063220661248 & 3.87 & 3.87 & 120.97 & -0.07 & 2 & 1 & 0.08 & 7.21 & 15.98 & 0 \\
2967553747040825856 & 3.80 & 3.80 & 330.27 & 0.07 & 0 & 1 & -0.00 & -66.11 & -10.92 & 0 \\
5948023586013872128 & 3.72 & 3.72 & 277.77 & 0.17 & 0 & 0 & 0.00 & -72.55 & -127.22 & 0 \\
5416221633076680704 & 3.65 & 3.65 & 126.59 & -0.08 & 0 & 0 & -0.01 & 69.68 & 19.93 & 0 \\
3222267297922229248 & 3.64 & 3.64 & 265.38 & -0.07 & 0 & 0 & 0.04 & -60.66 & 13.30 & 0 \\
6245775565362814976 & 3.59 & 3.59 & 133.82 & -0.08 & 0 & 0 & -0.07 & 194.28 & 59.86 & 0 \\
3235905365276381696 & 3.52 & 3.82 & 211.08 & -0.43 & 1 & 0 & 0.05 & -4.52 & 57.50 & 0 \\
3236272877038986240 & 3.47 & 3.47 & 141.67 & -0.06 & 0 & 0 & 0.06 & -50.01 & 10.89 & 0 \\
5200035927402217472 & 3.46 & 3.46 & 148.65 & -0.03 & 1 & 0 & 0.04 & -89.98 & 15.81 & 0 \\
5820283738165246976 & 3.45 & 3.45 & 380.58 & -0.08 & 0 & 0 & -0.02 & 55.83 & 78.12 & 0 \\
3222024374573501952 & 3.37 & 3.37 & 224.71 & -0.11 & 0 & 1 & 0.00 & -0.04 & 17.76 & 0 \\
3221019798902558720 & 3.37 & 3.37 & 142.94 & -0.07 & 0 & 0 & 0.03 & -64.74 & 48.73 & 0 \\
6235172592479759360 & 3.32 & 3.32 & 147.53 & 0.02 & 0 & 0 & 0.01 & -2.29 & 46.61 & 0 \\
3220688120051681792 & 3.27 & 3.27 & 131.65 & -0.11 & 0 & 0 & -0.01 & -11.03 & -126.68 & 0 \\
2920301135326368256 & 3.17 & 3.17 & 204.04 & -0.16 & 1 & 0 & 0.07 & -102.90 & 84.00 & 0 \\
6086395513764172800 & 3.11 & 3.11 & 130.44 & -0.16 & 0 & 0 & 0.04 & -55.69 & 27.51 & 0 \\
6014331906769737728 & 3.11 & 3.11 & 133.08 & -0.07 & 0 & 1 & -0.02 & 80.91 & 44.02 & 0 \\
3221921742035204608 & 3.09 & 3.10 & 114.30 & -0.02 & 0 & 0 & 0.02 & -46.60 & 11.19 & 0 \\
3181196759054802432 & 3.07 & 3.07 & 111.78 & -0.05 & 0 & 0 & -0.02 & 32.48 & 20.05 & 0 \\
6084509542084855808 & 3.03 & 3.03 & 191.97 & -0.03 & 0 & 0 & 0.00 & 25.25 & 92.65 & 0 \\
3217773074208416768 & 3.00 & 3.30 & 343.16 & -0.76 & 0 & 0 & -0.01 & -128.82 & 3.13 & 0 \\
6244153510833938432 & 3.00 & 3.00 & 117.06 & -0.08 & 0 & 0 & 0.01 & 41.17 & 35.66 & 0 \\
3220786839876234752 & 2.99 & 2.99 & 152.48 & -0.03 & 1 & 0 & 0.02 & -79.85 & -69.71 & 0 \\
\hline
	\end{tabular}
\end{table*}

%%%%%%%%%%%%%%%%%%%%%%%%%%%%%%%%%%%%%%%%%%%%%%%%%%

\begin{figure*}
	\centering
	\includegraphics[width=\textwidth]{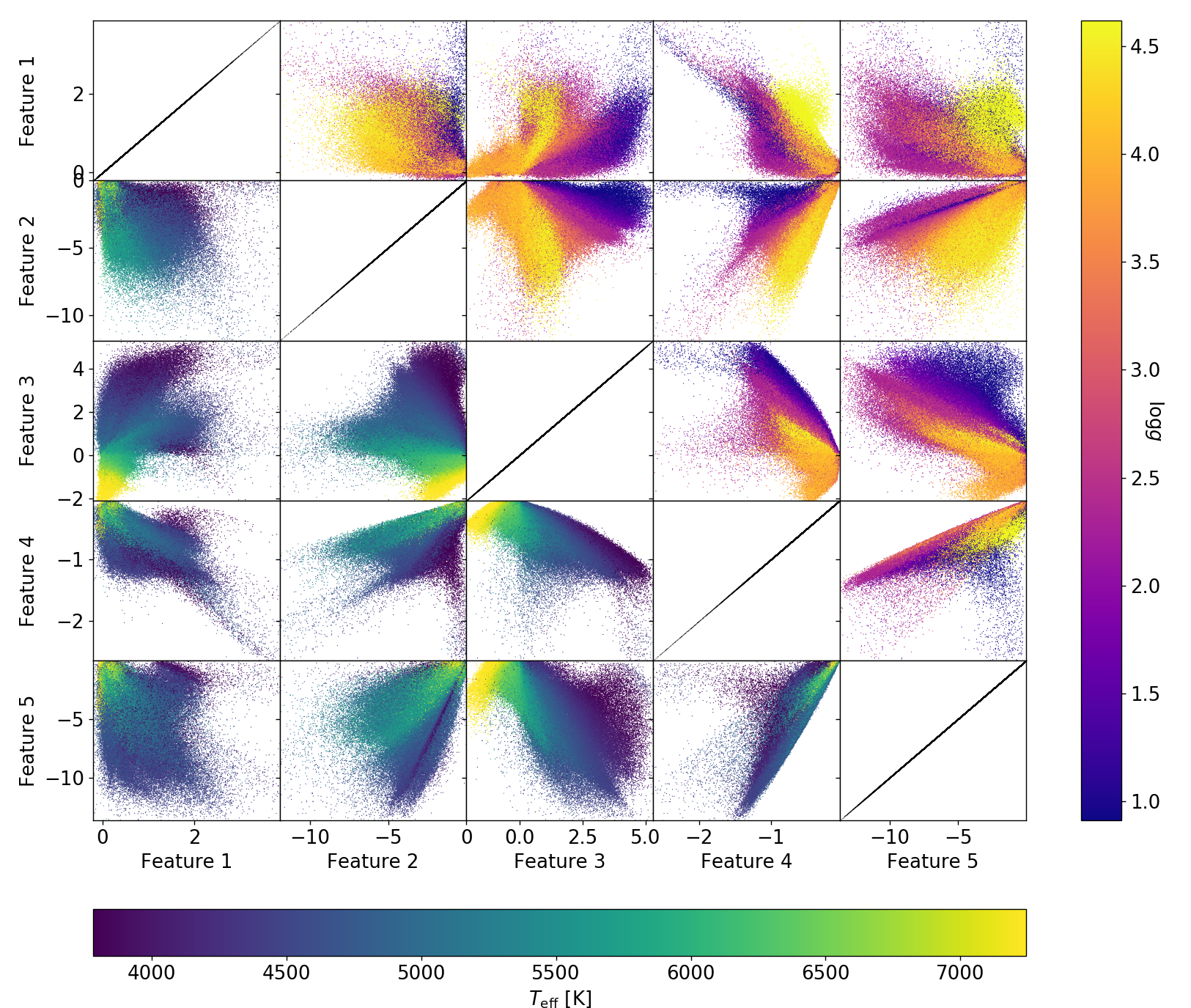}
	\caption{Same plots as shown in Figure \ref{fig:latent_ccd3_1}, but for the latent features of the blue HERMES band, coloured by parameter \Teff\ on lower triangle and by \Logg\ on the upper triangle.}
	\label{fig:latent_ccd1_1}
\end{figure*}

\begin{figure*}
	\centering
	\includegraphics[width=\textwidth]{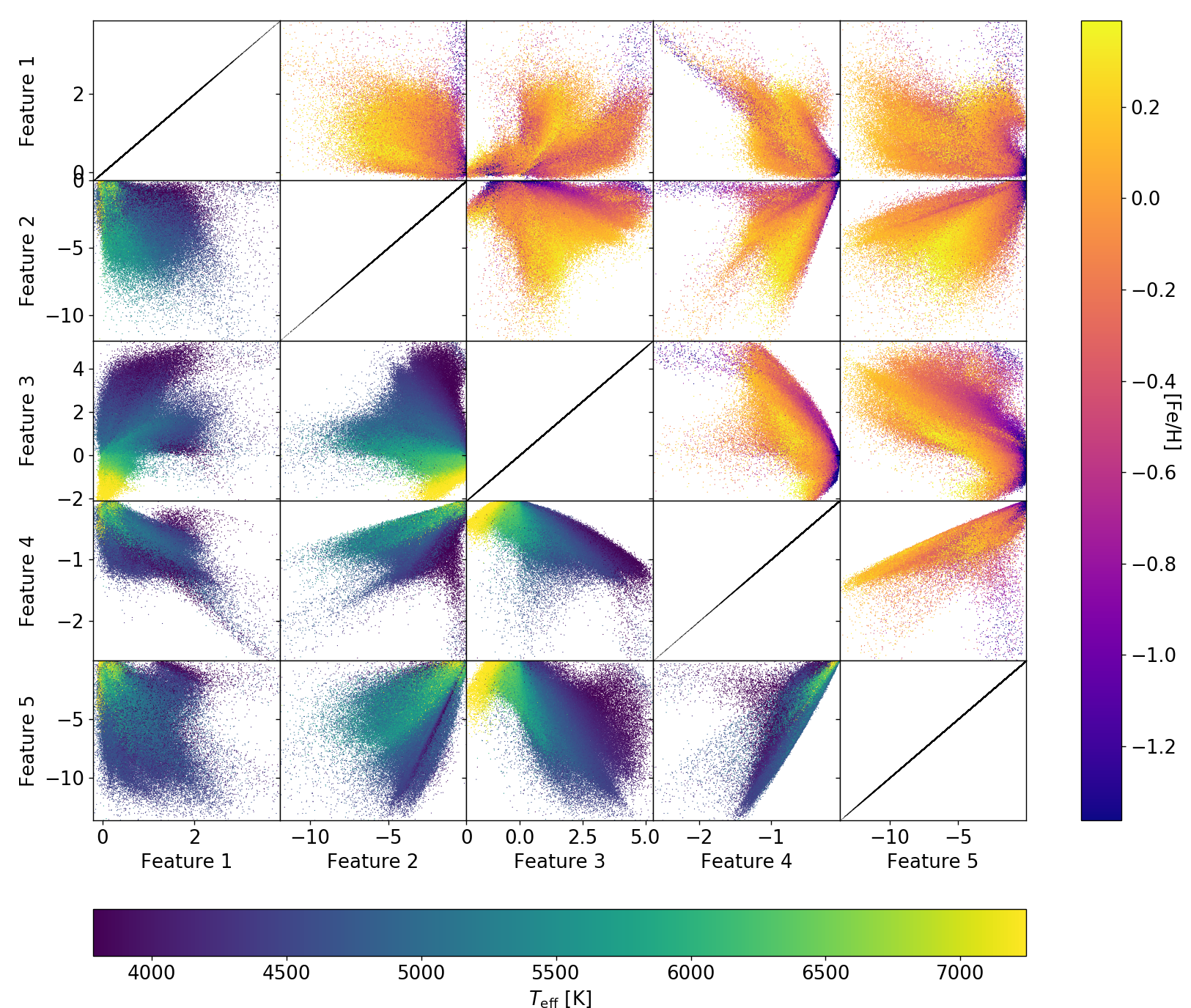}
	\caption{Same plots as shown in Figure \ref{fig:latent_ccd3_1}, but for the latent features of the blue HERMES band, coloured by parameter \Teff\ on lower triangle and by \Feh\ on the upper triangle.}
	\label{fig:latent_ccd1_2}
\end{figure*}

\begin{figure*}
	\centering
	\includegraphics[width=\textwidth]{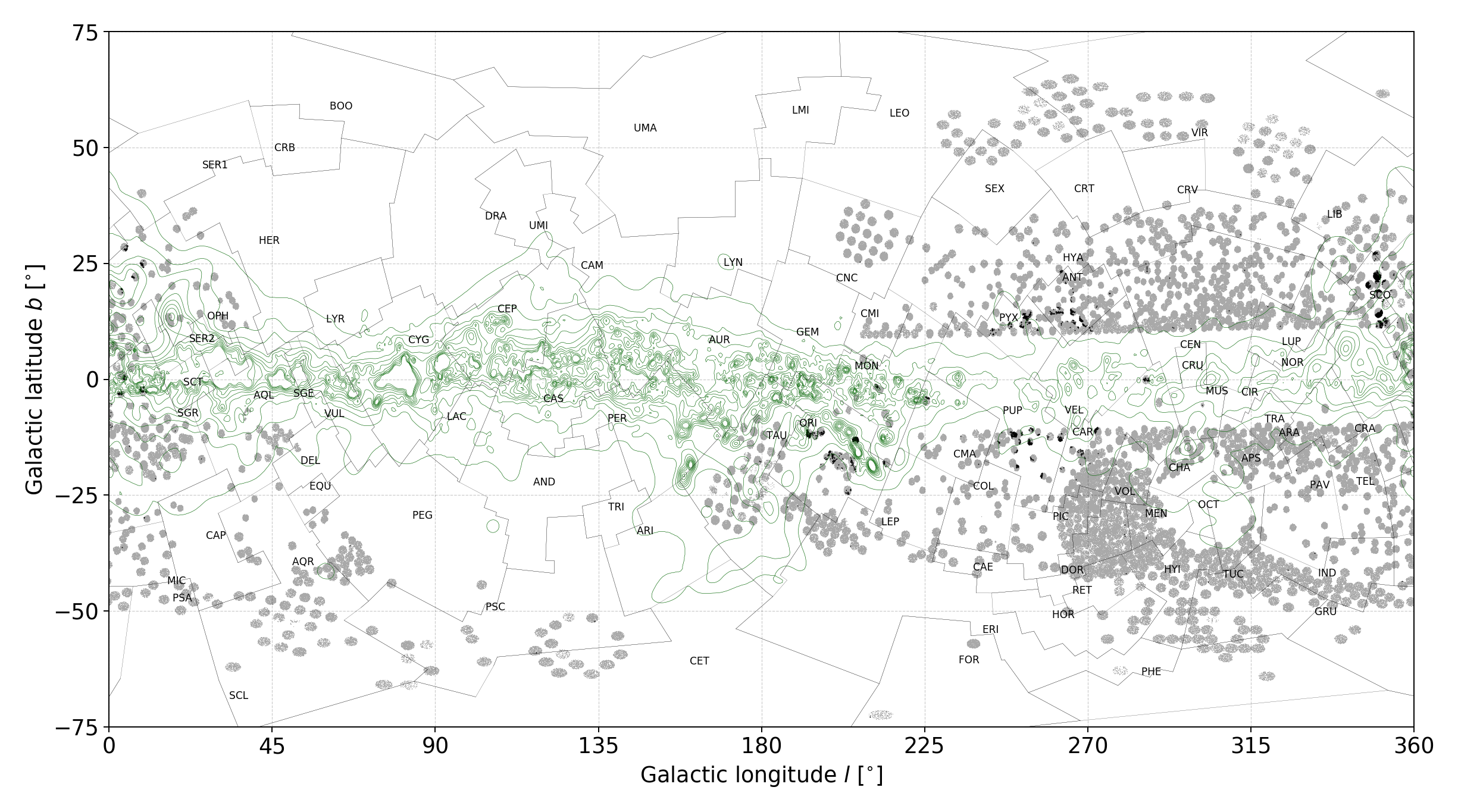}
	\caption{Same as Figure \ref{fig:spatialnebular} but showing stars with at least three detected nebula emission lines, shown with black dots.}
	\label{fig:spatialnebular}
\end{figure*}

%%%%%%%%%%%%%%%%%%%%%%%%%%%%%%%%%%%%%%%%%%%%%%%%%%

% Don't change these lines
\bsp	% typesetting comment
\label{lastpage}
\end{document}